\documentclass[a4paper,11pt]{article}
\pdfoutput=1 

\usepackage{jcappub} 

\usepackage{lscape}
\usepackage{verbatim}
\usepackage{comment}
\usepackage{array}
\usepackage{subcaption}
\usepackage{multirow}
\usepackage{graphicx}
\usepackage{amsmath}
\usepackage{amssymb}
\usepackage{rotating} 
\usepackage{xspace}
\usepackage{color}
\usepackage[x11names]{xcolor}
\usepackage{booktabs}
\usepackage{xcolor}
\usepackage[normalem]{ulem}
\usepackage{caption}
\usepackage[section]{placeins}

\DeclareMathOperator\erf{erf}

\newcommand{\Msun}{{\rm M}_\odot}

\title{The local dark matter distribution in self-interacting dark matter halos}

\author[a]{Elham Rahimi,}
\author[b]{Evan Vienneau,}
\author[b, c]{Nassim Bozorgnia,}
\author[d]{and Andrew Robertson}

\affiliation[a]{Department of Physics and Astronomy, York University,\\
4700 Keele Street, Toronto, Ontario M3J 1P3, Canada}
\affiliation[b]{Department of Physics, University of Alberta, \\
Edmonton, Alberta T6G 2E1, Canada}
\affiliation[c]{Theoretical Physics Institute, University of Alberta, \\
Edmonton, Alberta T6G 2E1, Canada}
\affiliation[d]{Jet Propulsion Laboratory, California Institute of Technology, \\
4800 Oak Grove Drive, Pasadena, CA 91109, USA}

\abstract{We study the effects of dark matter self-interactions on the local dark matter distribution in selected Milky Way-like galaxies in the EAGLE hydrodynamical simulations. The simulations were run with two different self-interacting dark matter models, a constant and velocity-dependent self-interaction cross-section. 
We find that the local dark matter velocity distribution of the Milky Way-like halos in the  simulations with dark matter self-interactions and baryons  are  generally similar to those extracted from cold collisionless dark matter simulations with baryons. In both cases, the local dark matter speed distributions agree well with their best fit Maxwellian distributions. Including baryons in the simulations with or without dark matter self-interactions increases the local dark matter density and shifts the dark matter speed distributions to higher speeds. To study the implications for direct detection, we compute the dark matter halo integrals obtained directly from the simulations and compare them to those obtained from the best fit Maxwellian velocity distribution. We find that a Maxwellian distribution provides a good fit to the halo integrals of most halos, without any significant difference between the results of different dark matter self-interaction models.  
}

\begin{document}
\maketitle

\section{Introduction}

The existence of dark matter (DM) is our current best explanation for a wide range of astronomical observations from galaxy rotation curves, gravitational lensing, hot gas in clusters, and the cosmic microwave background radiation. However, all evidence for DM originates from its gravitational interactions. The particle nature of DM and its non-gravitational interactions remain unknown. The current standard model of cosmology, $\Lambda$-cold dark matter ($\Lambda$CDM), assumes that DM is cold (non-relativistic at the epoch of matter-radiation equality) and collisionless (negligible self-interactions). Cosmological $\Lambda$CDM simulations yield excellent agreement with observations of the large-scale structure of the Universe \cite{Vogelsberger:2014kha, Schaye:2015}. However on small scales there exists some discrepancies.

Many of the classic challenges to $\Lambda$CDM on small scales such as the missing satellites \cite{Springel:2008cc, Bullock:2010uy, Fielder:2018szt}, core-cusp \cite{Navarro:2008kc, deBlok:2009sp, Kendall:2019fep} and too-big-to-fail \cite{Boylan-Kolchin:2011qkt, Garrison-Kimmel:2014} problems originated from a comparison of observations of dwarf galaxies with DM-only $\Lambda$CDM simulations~\cite{Spergel:2000}. It has since been shown that including baryons in the simulations can alleviate some of these tensions due to the effects of supernovae feedback and enhanced tidal stripping \cite{Brooks:2014, Sawala:2015cdf, Zhu:2015jwa, Brooks:2017}. However, additional challenges to $\Lambda$CDM on small scales including the diversity of rotation curves and planes of satellites problems \cite{Navarro:1997, Bullock2001, Oman2015} still remain. As a result, many alternative DM models have been proposed \cite{Goetz:2002vm, Viel:2013fqw, Newton:2020cog, Hu:2000, Du:2016zcv}.

Self-interacting DM (SIDM) is an alternative to collisionless CDM, in which DM particles have a non zero self-interaction cross section and scatter with one another at rates which are astrophysically significant \cite{Carlson1992, Vogelsberger2013, Robertson:2020pxj}. 
SIDM particle candidates are readily predicted by `hidden sector' particle physics models and can potentially resolve the $\Lambda$CDM small scale challenges \cite{Brooks:2012vi, Sawala:2015cdf,2016MNRAS.458.1559Z, 2017ApJ...850...97B,PhysRevLett.119.111102, Creasey_2017, PhysRevX.9.031020, Kahlhoefer_2019, PhysRevLett.124.141102, 2020PhRvL.124n1102S, Kaplinghat_2019, 2021MNRAS.503..920C} . 
Self-interactions transport heat from the hotter outer halo to the inner cooler halo leading to the transformation of higher density cusps into lower density cores \cite{Zavala:2013, Elbert:2015}. Furthermore, SIDM subhalos are less concentrated than their CDM counterparts and are therefore more prone to tidal disruption, leading to a reduced subhalo mass function \cite{Rocha:2013, Tulin2018}.

If DM is  self-interacting, its DM density and velocity distribution in the Solar neighborhood could be different from the case of collisionless DM. This can in turn impact the interpretation of signals in DM direct detection experiments, which aim to detect the  nuclear or electronic recoils induced by the scattering of a DM particle from the halo of the Milky Way (MW) in a low background underground detector. The interaction rate of DM particles with the target nucleus or electron in the detector depends on the DM density and velocity distribution in the Solar neighborhood. The DM distribution in our galaxy, on the other hand, strongly depends on the particle nature of DM and its interactions.

The simplest and most commonly adopted DM halo model is the Standard Halo Model (SHM). It assumes an isothermal, spherical DM halo with an isotropic Maxwell-Boltzmann DM velocity distribution. The local DM density in the SHM is assumed to be 0.3 or 0.4~GeV/cm$^3$ and the local velocity distribution is assumed to be a Maxwellian distribution in the Galactic rest frame, with a peak speed equal to the local circular speed, $v_c$, usually set equal to 230 km$/$s, and truncated at an escape speed of 544~km/s from the Galaxy. 
Recent cosmological simulations of  $\Lambda$CDM including baryons have provided realistic predictions for the local DM distribution and its implications for DM direct detection~\cite{Poole-McKenzie:2020dbo, Bozorgnia:2019mjk, Bozorgnia:2017brl, Bozorgnia:2016ogo, Kelso:2016qqj, Sloane:2016kyi}. These studies find that a Maxwell-Boltzmann distribution provides a good fit to the local DM velocity distribution of MW-like halos in $\Lambda$CDM simulations. 

The effects of DM self-interactions on the local DM distribution have been studied in the Aquarius SIDM simulations~\cite{Vogelsberger2013}. These simulations, which do not include baryons were run with  different SIDM models, including three different constant cross sections, as well as two velocity-dependent cross sections. They find that the DM velocity distribution at the Solar circle is Maxwellian for SIDM models with cross sections larger than 1 cm$^2$\,g$^{-1}$. For smaller self-interacting cross sections, the velocity distribution is non-Maxwellian, while for the velocity-dependent models, the velocity distribution is close to those extracted from collision-less DM-only simulations. High resolution SIDM simulations including baryons have recently become available~\cite{Robertson:2009bh}. Yet, a study of the local DM distribution and its implications for DM direct detection using such simulations had not been performed and is the aim of this work.

In this paper, we study the local DM density and velocity distribution of MW-like galaxies in SIDM versions~\cite{Robertson:2020pxj, Robertson:2017mgj} of the EAGLE (`Evolution and Assembly of GaLaxies and their Environments') simulations \cite{Schaye:2015, Crain:2015poa}, and discuss their implications for DM direct detection. The simulations include baryons and are performed for two SIDM models; DM with a velocity-independent and isotropic cross section of 1 cm$^2$~g$^{-1}$ (which we denote as ``SIDM1''), and DM with a velocity-dependent and anisotropic cross section corresponding to scattering through a Yukawa potential (which we denote as ``vdSIDM''). For each model, the CDM (collisionless DM + baryons) and DM-only counterparts of the SIDM halos (denoted as ``SIDM1-only'' and ``vdSIDM-only'') are available, and we compare our results for all these cases.

The paper is structured as follows. In section \ref{simulations} we present the details of the  hydrodynamical simulations used in this work. In section \ref{selection} we discuss our criteria for selecting the  MW analogues in the SIDM simulations and our method for identifying their CDM and SIDM-only counterparts. In sections \ref{density section} and \ref{vel_dist} we present the local DM density and velocity distribution of the simulated MW-like halos. 
In section \ref{halo} we present the \emph{halo integrals}, which together with the local DM density encapsulate the astrophysical dependence of direct detection event rates. 
Finally, in section \ref{conclusions} we discuss and summarize our main conclusions. In appendices~\ref{App.RC}, \ref{app:CDM}, \ref{app:Gen Maxwellian}, and \ref{app:parameters}, we provide additional materials relevant to this work.

\section{Simulations}\label{simulations}

In this work we use hydrodynamical simulations of MW-mass halos from the SIDM versions~\cite{Robertson:2020pxj, Robertson:2017mgj} of the EAGLE project~\cite{Schaye:2015, Crain:2015poa}. EAGLE is a suite of large-volume, cosmological simulations that employ state-of-the-art numerical techniques and subgrid models to simulate a vast array of astrophysical processes that are critical for galaxy formation.  The simulations use the SPH code, GADGET-3 \cite{Springel:2005mi} with modifications to the SPH, time-stepping and subgrid physics. The galaxy formation model used includes radiative cooling, star formation, stellar evolution, feedback due to stellar winds and supernovae, and the seeding, growth and feedback from black holes \cite{Crain:2015poa}. The subgrid parameters were calibrated to reproduce the galaxy stellar mass function at $z$ = 0.1 and the galaxy stellar mass vs.~size relation for disk galaxies.

We select candidate MW-like halos from a (50~Mpc)$^3$ simulation volume (denoted as ``EAGLE-50"), which has the same resolution level and galaxy formation model as the `Reference' (100 Mpc)$^3$ EAGLE box. The simulations have an initial gas particle mass, $m_{\rm gas}=1.8 \times 10^6~\Msun$, DM particle mass, $m_{\rm DM}=9.7 \times 10^6~\Msun$ (for SIDM and CDM) and $m_{\rm DM-only}=1.2 \times 10^7~\Msun$ (for DM-only), and a Plummer-equivalent gravitational softening length, $\epsilon_p=0.7$~kpc. They adopt the Planck-2013~\cite{Planck:2013pxb} cosmological parameters: $\Omega_m$ = 0.307, $\Omega_\Lambda$ = 0.693, $\Omega_b$ = 0.04825, $h = 0.6777$, $\sigma_8$ = 0.8288, and $n_s$ = 0.9611.

We investigate MW-like halos from SIDM1 and vdSIDM simulations along with their simulated CDM (collisionless DM + baryons) and  SIDM-only (including self-interactions, but no baryons) counterparts. The self-interaction of DM particles is implemented using the method introduced in ref.~\cite{Robertson:2016xjh}. At each time step, DM particles have some probability of scattering off a neighbour. The probability depends on the relative velocity and an azimuthally symmetric cross-section defined in the centre of momentum frame of the two particles. For SIDM1, the cross-section is velocity-independent, isotropic and equal to 1\, cm$^2$\,g$^{-1}$. For vdSIDM, the cross-section is velocity-dependent and anisotropic, corresponding to DM particles scattering through a Yukawa potential. The differential cross section which is used for vdSIDM in the simulations is given by~\cite{Robertson:2020pxj},
\begin{equation}
\frac{d\sigma}{d\Omega}=\frac{\sigma_{T0}}{4\pi\left[1+\left(v^2/w^2\right)\sin^2(\theta/2)\right]^2},    
\end{equation}
where $\sigma_{T0}/m=3.04$~cm$^2$\,g$^{-1}$, with $m$ the DM particle mass, and $w=560$~km\,s$^{-1}$. The parameters of this model were chosen such that the resulting DM density profiles would fit the observations from dwarf galaxies to galaxy clusters~\cite{Robertson:2020pxj, Kaplinghat:2015aga}.

\section{Selection of Milky Way-like galaxies}\label{selection}

To make accurate predictions for the local DM distribution in the MW, we need to select simulated halos which resemble the MW galaxy. We identify simulated \emph{MW-like} galaxies according to the following criteria: (i) their virial\footnote{Virial quantities are defined  as those corresponding to a sphere centered on the galaxy's minimum of the potential, containing a mean density 200 times the critical density.} mass, $M_{200}$, is in the range $[0.5 - 3] \times 10^{12}$ $\Msun$~\cite{Callingham:2018vcf}, (ii)  their stellar mass lies within the 3$\sigma$ range of the observed MW stellar mass, [4.5 - 8.3] $\times 10^{10}~\Msun$~\cite{McMillan:2011wd}, (iii) their rotation curve provides a good fit to recent observations of the MW rotation curve from ref.~\cite{Iocco:2015xga}, and (iv) the halos are relaxed. 

For criterion (iii), we compare the rotation curves of the simulated halos to the large compilation of observational measurements of the MW rotation curve presented in ref.~\cite{Iocco:2015xga}. For each simulated galaxy, we compute the circular velocity, $v_c(r) = \sqrt{GM(<r)/r}$, by determining the total mass $M(<r)$ (in gas, stars, black holes and DM) enclosed within a sphere of Galactocentric radius $r$. We then compute the angular circular velocity, $\omega_c(r)=v_c(r)/r$, and follow the procedure given in ref.~\cite{Calore:2015oya} and discussed further in appendix \ref{App.RC}, to find the goodness-of-fit to the observed rotation curve data. 
As detailed in ref.~\cite{Iocco:2015xga}, using the angular circular velocity for fitting purposes is more convenient than working with the circular velocity. This is because the errors in $\omega_c(r)$ and $r$ are not correlated, while errors in $v_c(r)$ and $r$ are correlated. To find the goodness-of-fit, we only consider Galactocentric distances greater than 2.5 kpc, since the orbits of tracers in ref.~\cite{Iocco:2015xga} become non-circular at smaller radii due to the effect of the galactic bulge.

For criterion (iv), we define relaxed halos according to the definitions in ref.~\cite{Neto:2007vq}. In particular, halos are defined as relaxed if they have a mass fraction in resolved subhalos within the virial radius, $f_{\rm sub} < 0.1$, the separation between the most bound particle and the centre of mass of the halo,  $s < 0.07$ (in units of the virial radius), and  a \emph{virial-ratio}, $2T/|U|<1.35$, where $T$ is the kinetic energy of the halo particles within the virial radius, including thermal energy of gas, and $U$ is the gravitational potential.

With criteria (i) and (ii) only, we identify 23 SIDM1 and 31 vdSIDM MW-like halos. Adding criterion (iii), we further reduce the number of MW-like halos to 16 SIDM1 and 18 vdSIDM. Lastly, applying criterion (iv) selects out one unrelaxed SIDM1 halo and one unrelaxed CDM counterpart of an SIDM1 halo, which we remove from our selected SIDM1 halos. This yields a total of 14 SIDM1 MW-like halos. Criterion (iv) also selects out one unrelaxed vdSIDM halo, yielding a total of 17 vdSIDM MW-like halos. The stellar and virial masses of our selected MW-like halos in are given in table~\ref{table1}. 

\begin{table}[h]
\resizebox{0.5\columnwidth}{!}{
\renewcommand{\arraystretch}{1.205}
\begin{tabular}[]{ |c|c|c|c|  }
 \hline
 \multicolumn{3}{|c|}{SIDM1} \\
 \hline
 Halo Number &$M_{\star}$ [$\times 10^{10}~\Msun$]&$M_{200}$ [$\times 10^{12}~\Msun$]\\
 \hline
 48 &5.32 & 2.54 \\
 60 &7.33 & 2.79 \\
 71 &5.47 & 1.78 \\
 74 &5.77 & 2.95 \\
 80 &5.75 & 2.70 \\
 92 &6.08 & 2.26 \\
 96 &5.82 & 2.35 \\
 99 &6.55 & 2.24 \\
 102 &5.03 & 1.88 \\
 111 &4.68 & 1.45 \\
 118 &5.08 & 1.79 \\
 128 &5.35 & 1.65 \\
 131 &5.07 & 1.62 \\
 159 &4.79 & 1.33 \\
 \hline
\end{tabular}%
}
\renewcommand{\arraystretch}{1}
\resizebox{0.5\columnwidth}{!}{%
\begin{tabular}[]{ |c|c|c|c|  }
 \hline
 \multicolumn{3}{|c|}{vdSIDM} \\
 \hline
 Halo Number &$M_{\star}$ [$\times 10^{10}~\Msun$]&$M_{200}$ [$\times 10^{12}~\Msun$] \\
 \hline
 55 &6.02 & 2.61 \\
 60 &7.13 & 2.85 \\
 72 &5.08 & 1.72 \\
 74 &5.75 & 2.91 \\
 75 &6.56 & 2.81 \\
 77 &4.70 & 2.30 \\
 92 &5.70 & 2.55 \\
 93 &7.23 & 2.30 \\
 99 &5.42 & 1.93 \\
 101 &5.18 & 2.45 \\
 107 &4.52 & 1.59 \\
 115 &4.69 & 1.43 \\
 122 &4.59 & 1.81 \\
 124 &5.14 & 1.61 \\
 128 &4.69 & 1.43 \\
 138 &4.73 & 1.58 \\
 176 &4.55 & 1.30 \\
 \hline
\end{tabular}%
}
\captionof{table}{\label{table1} Halo number, stellar mass within 30 kpc of the Galactic center, $M_\star$, and virial mass, $M_{\rm 200}$, for the selected MW-like halos in the EAGLE-50 SIDM1 (left table) and vdSIDM (right table) simulations.}
\end{table}

Finally, for each MW-like halo satisfying our criteria, we select its CDM and SIDM-only counterparts by identifying the halo that minimizes the 4D Euclidean distance defined as $\sqrt{[\Delta r/1~{\rm kpc}]^2 + [\Delta\log(M_{200}/\Msun)/0.1]^2}$, where $\Delta r$ is the 3D Euclidean distance between the two halo positions in their respective simulation boxes and $\Delta \log(M_{200}/\Msun)$ is the difference between the logarithm of the virial masses of the two halos. Notice that we do not impose our MW-like galaxy selection criteria on the CDM counterparts. In appendix~\ref{app:CDM} we explore how the results change for the CDM halos in EAGLE-50 which satisfy our MW-like galaxy selection criteria.

\begin{figure}[t]\label{rotation_curve_plot}
\centering
    \includegraphics[width=\textwidth]{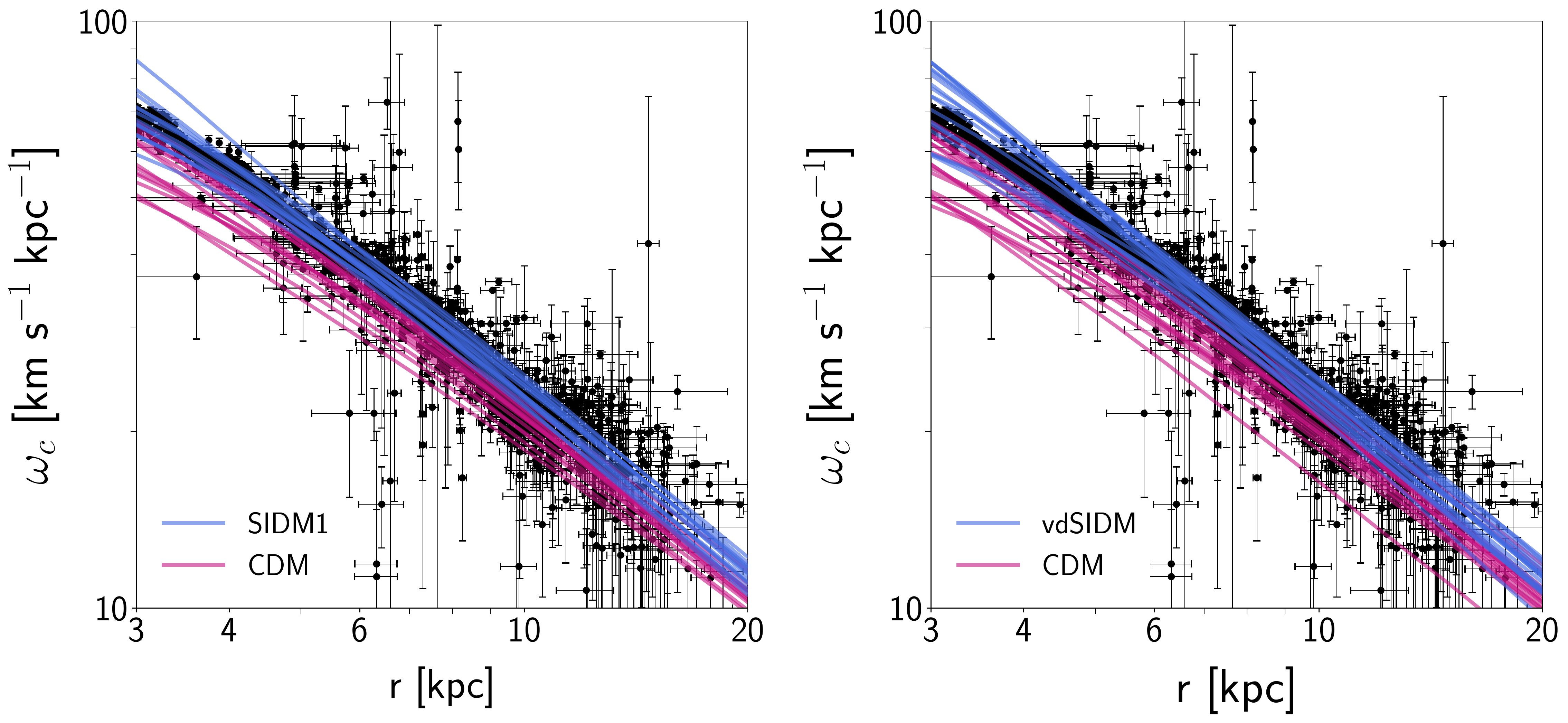}
  \caption{Angular circular velocity, $\omega_c$, as a function of the Galactocentric distance, $r$, for the selected EAGLE-50 SIDM1 (left panel) and vdSIDM (right panel) MW-like halos (blue curves) and their corresponding CDM counterparts (magenta curves). The black points show observational measurements of the MW rotation curve from ref.~\cite{Iocco:2015xga} and the error bars represent the $1\sigma$ errors.}
  \label{fig:rotation curves}
\end{figure}

In figure~\ref{fig:rotation curves}, we show the angular circular velocity, $\omega_c$, as a function of the Galactocentric distance, $r$, for our selected SIDM1 (left panel) and vdSIDM (right panel) halos as blue curves, with their corresponding CDM counterparts shown as magenta curves. 
The data from observational measurements of the MW rotation curve  is shown as black points with their $1\sigma$ error bars. We can see from the figure that  the rotation curves of the selected SIDM halos in EAGLE-50 agree well with the observed MW rotation curve, while the rotation curves of their CDM counterparts are slightly shifted to lower angular circular velocities.

\section{Local dark matter density}\label{density section}

The DM density in the Solar neighborhood enters as a normalization in the event rate calculation for direct detection experiments. The fiducial value adopted for the local DM density in the SHM is 0.3~GeV$/$cm$^3$. 
In this section we present the local DM density and the maximum DM density variation throughout the Solar neighbourhood for our selected MW-like halos in EAGLE-50.

The particle positions and velocities for each simulated halo are given with respect to the simulation reference frame. To transform them to the Galactic reference frame, we subtract the particle positions by the center of potential, which is the position of the most gravitationally bound particle in the halo. The particle velocities are also transformed, such that they are with respect to the halo center. We align the $z$-axis with the average angular momentum of the star particles within 10 kpc from the halo center. We then define the $x$ and $y$ axes to lie in the plane perpendicular to the $z$-axis, which is the plane of the stellar disk. We adopt the same coordinate transformation for  the SIDM-only halos as their SIDM counterparts.

To determine the {\it local} DM density, we consider the density of DM particles in a torus region aligned with the stellar disk, with inner and outer Galactocentric radii of 7 kpc and 9 kpc, and height between $-2$~kpc and 2 kpc with respect to the Galactic plane. In the case of the SIDM-only halos, we orientate our torus to match the orientation of their SIDM counterpart. The number of DM particles in the torus region in the Solar neighbourhood, the average DM density in the torus and the maximum DM density variation within the torus for  MW-like halos in each SIDM model, as well as for their SIDM-only and CDM counterparts are given in table~\ref{table3_density}. The range of numbers presented in the table correspond to the halo-to-halo variation for each case. The Poisson uncertainty in the local DM densities is on average at the $\sim 10$\% level.

\begin{table}[t]
\centering
\begin{tabular}[]{ |c|c|c|c|  }
 \hline
   & DM particles & Local DM density [GeV/cm$^3]$ & DM density variation $\%$ \\
 \hline
 SIDM1 & $447 - 717$ & $0.41 - 0.66$  & $4 - 26$  \\
SIDM1-only & $274 - 544$  & $0.30 - 0.59$  &  $4 - 53$ \\
CDM &  $380 - 729$ & $0.35 - 0.67$  & $4 - 41$ \\
 \hline
  vdSIDM & $325 - 734$  & $0.30 - 0.67$  & $5 - 39$ \\
vdSIDM-only & $216 - 496$ & $0.23 - 0.54$  & $15 - 54$\\
CDM & $373 - 729$ & $0.34 - 0.67$  &$4 - 41$\\
 \hline
 \end{tabular}
\captionof{table}{\label{table3_density} The number of DM particles in the torus region, the local DM density and Poisson uncertainty (in units of GeV/cm$^3$), and the percentage of the local DM density variation for the selected SIDM1 and vdSIDM1 halos, as well as their SIDM-only and CDM counterparts in the EAGLE-50 simulations. 
}
\label{table of mw reduced chi squared highres}
\end{table}

Table \ref{table3_density} shows that the local DM densities of the SIDM1 and vdSIDM halos agree well with their CDM counterparts, and are almost always larger than the fiducial value of 0.3~GeV$/$cm$^3$. For the SIDM-only halos, the local DM density values are systematically smaller than their SIDM and CDM counterparts. The increase in the local DM density in hydrodynamic simulations compared to SIDM-only is a result of the  contraction of the DM halo due to dissipational baryonic processes. 
In general, the local DM densities of the SIDM1, vdSIDM, and their CDM counterparts agree well with global~\cite{McMillan:2011wd, Catena:2009mf, Weber:2009pt, Iocco:2011jz, Nesti:2013uwa, Sofue:2015xpa, Pato:2015dua, deSalas:2019pee} and local~\cite{Salucci:2010qr, Smith:2011fs, Bovy:2012tw, Garbari:2012ff, Zhang:2012rsb, Bovy:2013raa, 2018A&A...615A..99H, Buch:2018qdr} estimates from observations, as well as those obtained previously for CDM halos in the EAGLE and APOSTLE~\cite{Bozorgnia:2016ogo}, and Auriga~\cite{Bozorgnia:2019mjk} simulations.

The DM density variation throughout the Solar neighbourhood is computed by dividing the torus into four  equally sized regions for each halo and computing the percent difference between the lowest and highest density region. The number of equally sized regions is chosen such that there is a statistically significant number of particles in each region. As shown in table~\ref{table3_density}, the maximum DM density variation reaches $\sim 40\%$ for the vdSIDM halos, which is smaller than the halo-to-halo variation in the local DM density.

\section{Dark matter velocity distributions}\label{vel_dist}

The local DM velocity  distribution is an important astrophysical input in the calculation of direct detection event rates. In this section we discuss and compare the DM velocity distribution in the Solar neighborhood for the SIDM1 and vdSIDM halos, and their SIDM-only and CDM counterparts. In section~\ref{velmod}, we discuss the local DM velocity modulus distribution, and in section~\ref{velcomp} we present the vertical, radial and azimuthal components of the DM velocity distribution.

\subsection{Velocity modulus distribution}\label{velmod}

The  DM velocity modulus distribution, $f(v)$, is related to the velocity distribution $\tilde{f}(\bf v)$ by,
\begin{equation}
    f(v) = v^2\int d\Omega_{\mathbf{v}}\tilde{f}(\mathbf{v}),
\end{equation}
where $d\Omega_{\bf v}$ is an infinitesimal solid angle around the direction $\mathbf{v}$. The velocity modulus distribution is normalized as $\int dv f(v) = 1$, such that $\int d^3 v \tilde{f}({\bf v})=1$. We compute the local DM velocity modulus distribution of each selected halo in the Galactic reference frame by finding the average speed distribution of DM particles in the torus region in the Solar neighborhood.

\begin{figure}[t]
 \begin{center}
    \includegraphics[width=0.8\textwidth]{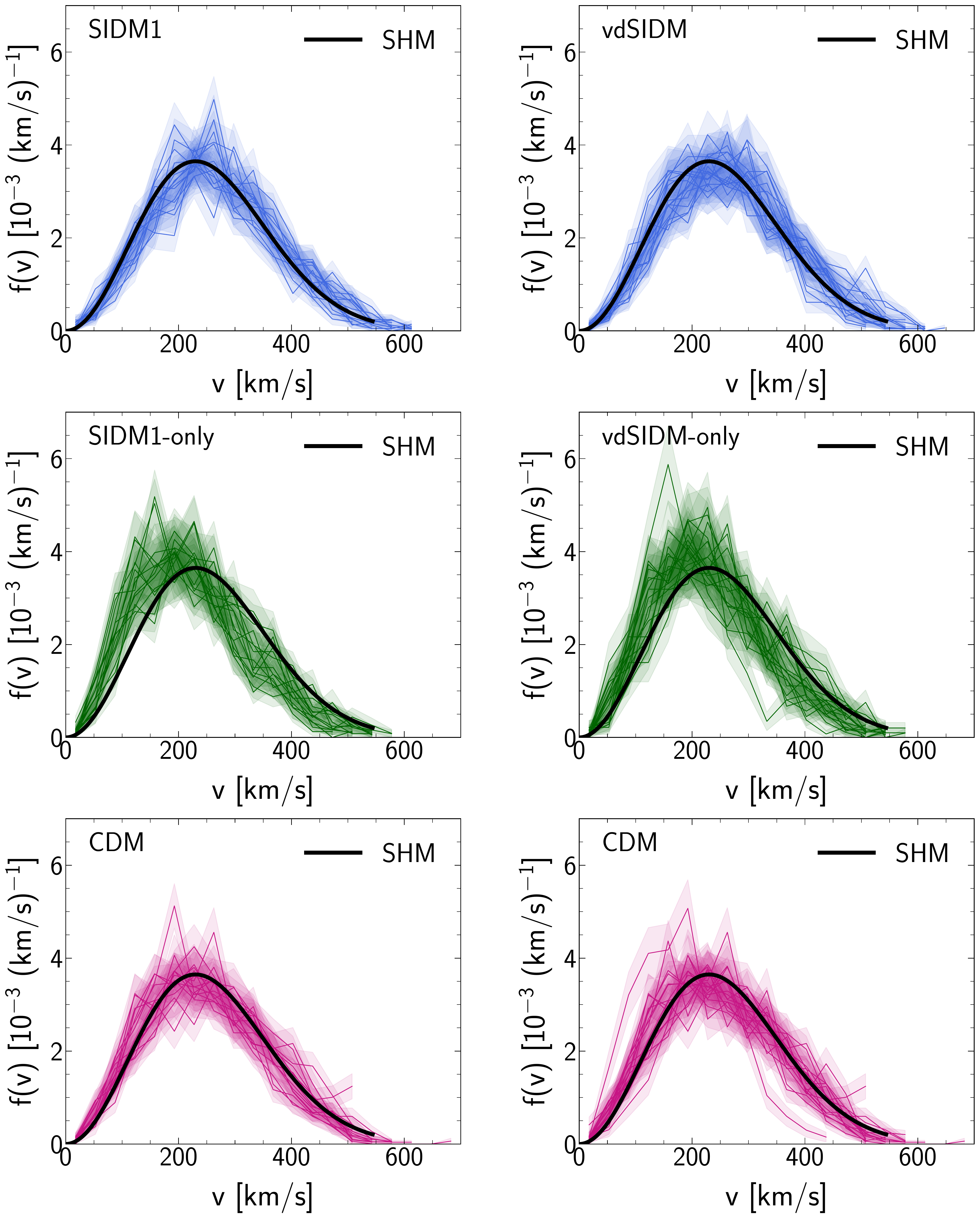}
   \end{center}
  \caption{The local DM velocity modulus distributions (solid coloured curves) and their 1$\sigma$ Poisson uncertainties (shaded bands) in the Galactic rest frame  for all selected  MW-like galaxies in the EAGLE-50  simulations. The top panels show the results for the SIDM1 and vdSIDM halos, while the middle and bottom panels show the results for their SIDM-only and CDM counterparts. The solid black curve specifies the SHM Maxwellian distribution with a peak speed of 230 km$/$s.}
  \label{all_vel_dist}
\end{figure}

Figure \ref{all_vel_dist} shows the local DM velocity modulus distributions (solid coloured lines) and their 1$\sigma$ Poisson uncertainties (shaded bands) in the Galactic reference frame for all selected MW-like galaxies in the EAGLE-50 simulations. The top panels show the results for the SIDM1 and vdSIDM halos, while the results for the corresponding SIDM-only and CDM counterparts are shown in the middle and bottom panels, respectively. The solid black line represents the SHM Maxwellian distribution with a peak speed of 230 km$/$s. We use a speed bin size of 35 km/s to compute the velocity modulus distribution, which is chosen to reduce noise while retaining the important features of the distributions.

As it can be seen from figure \ref{all_vel_dist}, we find no significant differences between the SIDM1 and vdSIDM speed distributions. We also see that the inclusion of baryons in the simulations leads to a shift of the peak speed of the distributions towards higher speeds. This is a result of the deepening of the gravitational potential well of the halo due to the presence of baryons, resulting in higher velocity particles. Including baryons along with DM self-interactions results in an even larger shift towards higher speeds. This can be inferred from the slightly larger peak speeds of the DM velocity distributions of the SIDM halos as compared to their CDM counterparts. This is due to the lower rotation curves of the CDM halos compared to their SIDM counterparts, as seen in figure~\ref{fig:rotation curves}, and the fact that the CDM counterparts are not all MW-like. In figure~\ref{fig: cdm_speed_and_halo} of appendix~\ref{app:CDM}, we present the local DM velocity modulus distributions of CDM halos in EAGLE-50 that satisfy our MW-like halo selection criteria, but are not all counterparts of the SIDM halos. For the MW-like CDM halos, the DM velocity modulus distributions are similar to SIDM halos, without showing the small shift towards higher speeds. 

Next we explore how well a Maxwell-Boltzmann distribution with a free peak speed fits the velocity distributions of the simulated halos. For each of our selected halos, we  fit the velocity modulus distribution with a  Maxwellian distribution truncated at the local Galactic escape speed, $v_{\rm esc}$, given by
\begin{equation}
f(v) =\begin{cases}
\frac{1}{N_{\rm esc}}\frac{4v^2}{\sqrt{\pi}v_0^3}\exp\left(-\frac{v^2}{v_0^2}\right)   , & {\rm for}~{v}<v_{\rm esc} \\
0, & {\rm otherwise}
\end{cases}
\label{eq:maxwellian}
\end{equation}
where
\begin{equation}
N_{\rm esc} = \erf(z) - 2z \exp{\left(-z^2\right)}/\pi^{1/2},    
\end{equation}
with $z=v_{\rm esc}/v_0$. For each halo, we determine $v_{\rm esc}$ by finding the DM particle in the torus with the highest speed. The escape speeds found in this way are in the range of $[519-625]$~km/s for the SIDM1 halos and $[490-650]$~km/s for the vdSIDM halos, which agree well with the measurements of the  MW escape speed~\cite{2018A&A...616L...9M, 2019MNRAS.485.3514D, Piffl:2013mla, Piffl:2014mfa}. 

\begin{figure}[t]
\begin{center}
    \includegraphics[width=0.8\textwidth]{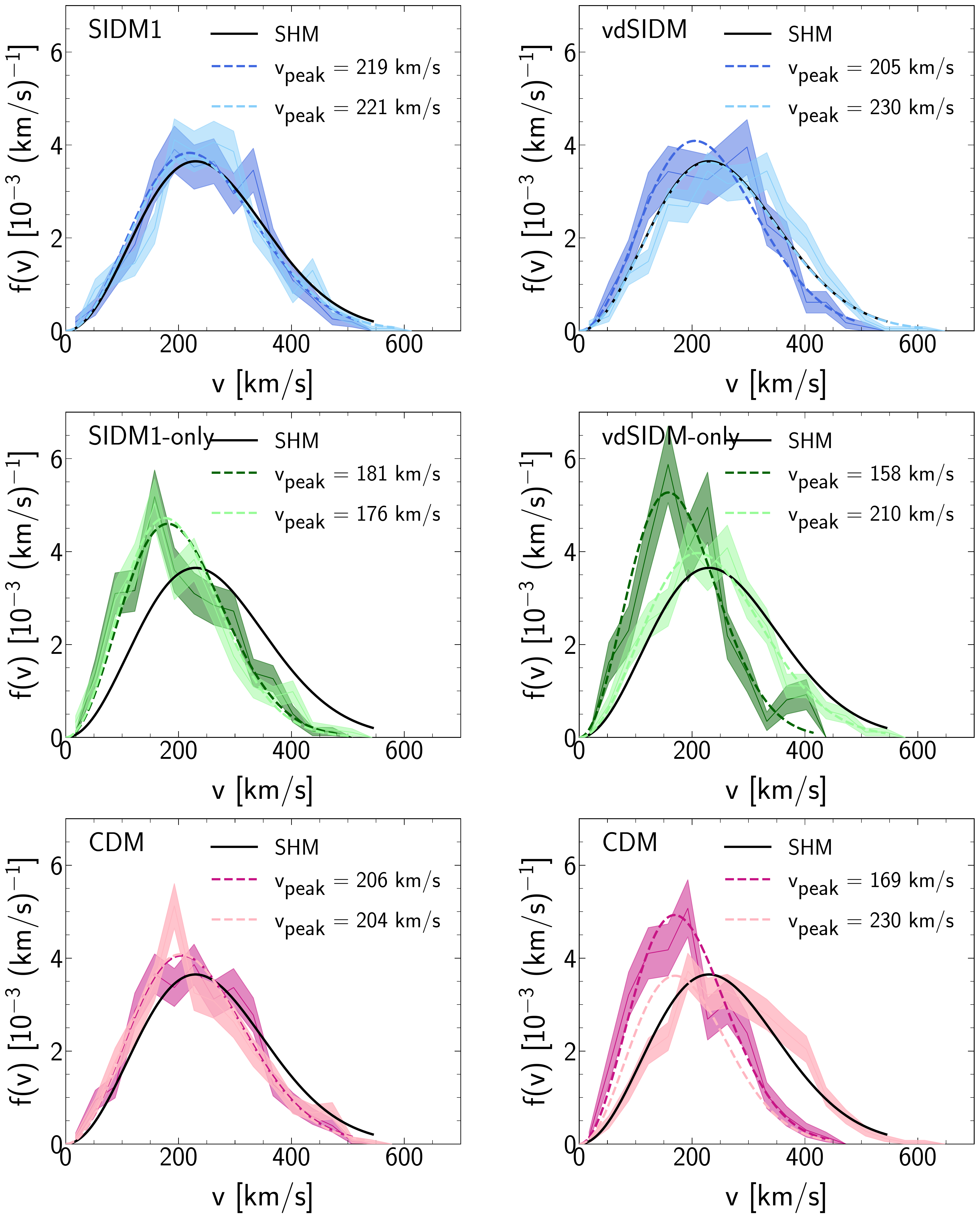}
\end{center}
  \caption{The local DM velocity modulus distributions (solid coloured curves) and their 1$\sigma$ Poisson uncertainties (shaded bands) in the Galactic rest frame for the SIDM1 and vdSIDM  halos (top panels) that have the best (dark colour) and worst (light colour) fit to the truncated Maxwellian velocity distribution. The middle and bottom panels  represent the SIDM-only and CDM counterparts, respectively. The best fit truncated  Maxwellian distribution for each halo is plotted as the dashed curve, and the SHM Maxwellian distribution with a peak speed of 230 km$/$s is shown as the black solid curve. For each halo the value of the best fit peak speed, $v_{\rm peak}$, of the Maxwellian distribution is indicated on the panels. }
  \label{fig: best_worst_vel_dist}
\end{figure}

Figure~\ref{fig: best_worst_vel_dist} shows the local DM velocity modulus distributions (solid coloured lines) and their 1$\sigma$ uncertainties (shaded bands) for the  SIDM1 and vdSIDM  halos (top panels) that have the best (dark colour) and worst (light colour) fit to the truncated Maxwellian velocity distribution. The corresponding SIDM-only and CDM counterparts are shown in the middle and bottom panels, respectively. For each halo, the dashed line of the same colour indicates the best fit Maxwellian speed distribution, with the value of the best fit peak speed also specified. In each panel, the SHM Maxwellian distribution with a peak speed of 230 km$/$s is shown with a solid black line.  

We test the goodness of fit of the truncated standard Maxwellian distribution to the local DM velocity modulus distributions of all selected SIDM1 and vdSIDM halos, as well as their SIDM-only and CDM counterparts. In appendix \ref{app:Gen Maxwellian}, we also explore the fits to a truncated generalized Maxwellian distribution. In tables~\ref{tab:eagle-50-sidm1}--\ref{tab:eagle-50-vdsidm-CDM} of appendix~\ref{app:parameters}, we list the best-fit parameters for both the Maxwellian and generalized Maxwellian distributions for each SIDM, SIDM-only, and CDM halo. In general, the standard Maxwellian distribution provides a good fit to DM velocity modulus distributions of the SIDM1, vdSIDM, CDM, SIDM1-only, and vdSIDM-only halos in the EAGLE-50 simulations. 
The truncated generalized Maxwellian distribution yields a slightly better fit to the local DM speed distributions for some  halos, but in general does not provide a significant improvement in the goodness of fits (see appendices \ref{app:Gen Maxwellian} and \ref{app:parameters}).

The range of the best fit peak speeds of the truncated standard Maxwellian distribution is $[218 - 246]$~km/s for the SIDM1 halos, while it is $[176 - 219]$~km/s and $[204 - 241]$~km/s for their corresponding SIDM1-only and CDM counterparts, respectively. For the vdSIDM halos, the range of the best fit peak speeds is $[205 - 258]$~km/s and the ranges for the corresponding vdSIDM-only and CDM halos are $[158 - 238]$~km/s and $[169 - 241]$~km$/$s, respectively. In each case, the SIDM and CDM halos have similar ranges of best fit peak speeds, while the ranges are smaller for the SIDM-only halos. This is due to shallower gravitational potential of the SIDM-only halos, shifting the DM speed distributions to smaller speeds.

\begin{figure}[t]
\centering
  \begin{subfigure}[b]{0.9\textwidth}
    \includegraphics[width=\textwidth]{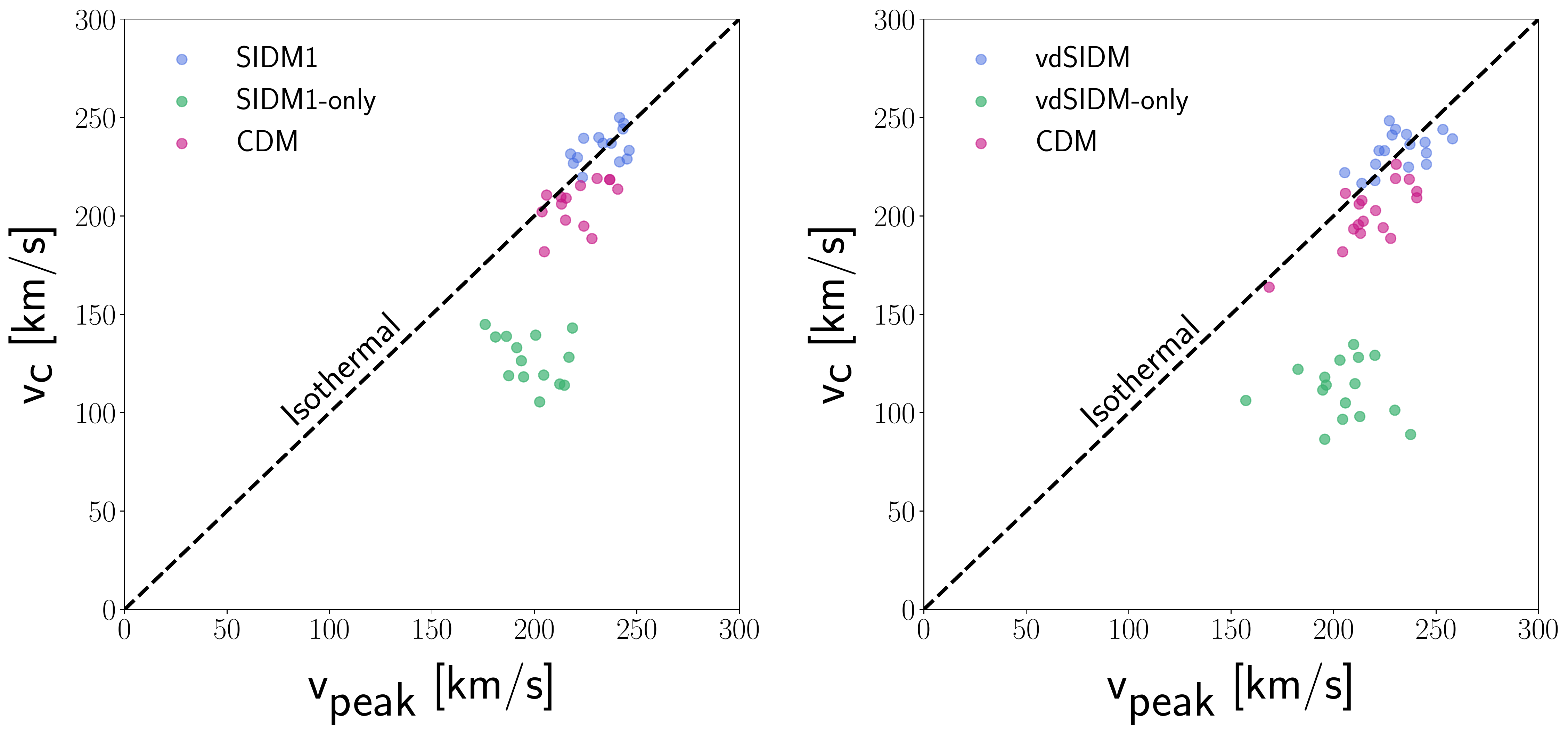}
  \end{subfigure}  
  \caption{Local circular speed at 8 kpc from the Galactic center, $v_c$, and the best fit peak speed, $v_{\rm peak}$, of the truncated Maxwellian distribution for the SIDM1 (left panel) and vdSIDM (right panel) halos in the EAGLE-50 simulations. In each panel, the SIDM-only (green dots) and CDM (magenta dots) counterparts of the SIDM (blue dots) halos are also shown. The dashed black line specifies the isothermal distribution with $v_c = v_{\rm peak}$.} 
  \label{peak_speed_circular_speed}
\end{figure}

To evaluate how close the SIDM halos are to an isothermal halo, in figure~\ref{peak_speed_circular_speed} we compare the local circular speed, $v_c$, and the best fit Maxwellian peak speed, $v_{\rm peak}$, of the SIDM1 (left panel) and vdSIDM (right panel) halos. The SIDM-only and CDM counterparts are also shown in each panel for comparison. The dashed lines indicate the isothermal distribution where the local circular speed equals the peak speed. It can be seen from the figure that the SIDM1 and vdSIDM halos along with their CDM counterparts are in good agreement with an isothermal distribution. Most of the CDM halos have smaller $v_c$ and $v_{\rm peak}$ values compared to their SIDM counterparts. This is again due to the fact that the CDM halos are not all MW-like, and have lower rotation curves compared to their SIDM counterparts, which results in smaller $v_c$ and $v_{\rm peak}$ values. 
The SIDM1-only and vdSIDM-only halos strongly deviate from an isothermal distribution, as expected.

\subsection{Velocity distribution components}\label{velcomp}

Next we study the velocity distribution components to determine whether any velocity anisotropies exist at the Solar radius. We present the components of the velocity distributions in the radial ($v_\rho$), azimuthal ($v_\phi$) and vertical ($v_z$) directions in the Galactic reference frame. 
We then fit each component of the velocity distributions with a Gaussian function given by
\begin{equation}
f(v_{i}) = \frac{1}{\sqrt{\pi}v_0} \exp{\left[-(v_i-\mu)^2/v_0^2\right]} \,,
\label{eq:gaussian}
\end{equation}
where $v_0$ and $\mu$ are free parameters.

\begin{figure}[h!]
  \centering
    \includegraphics[width=0.8\textwidth]{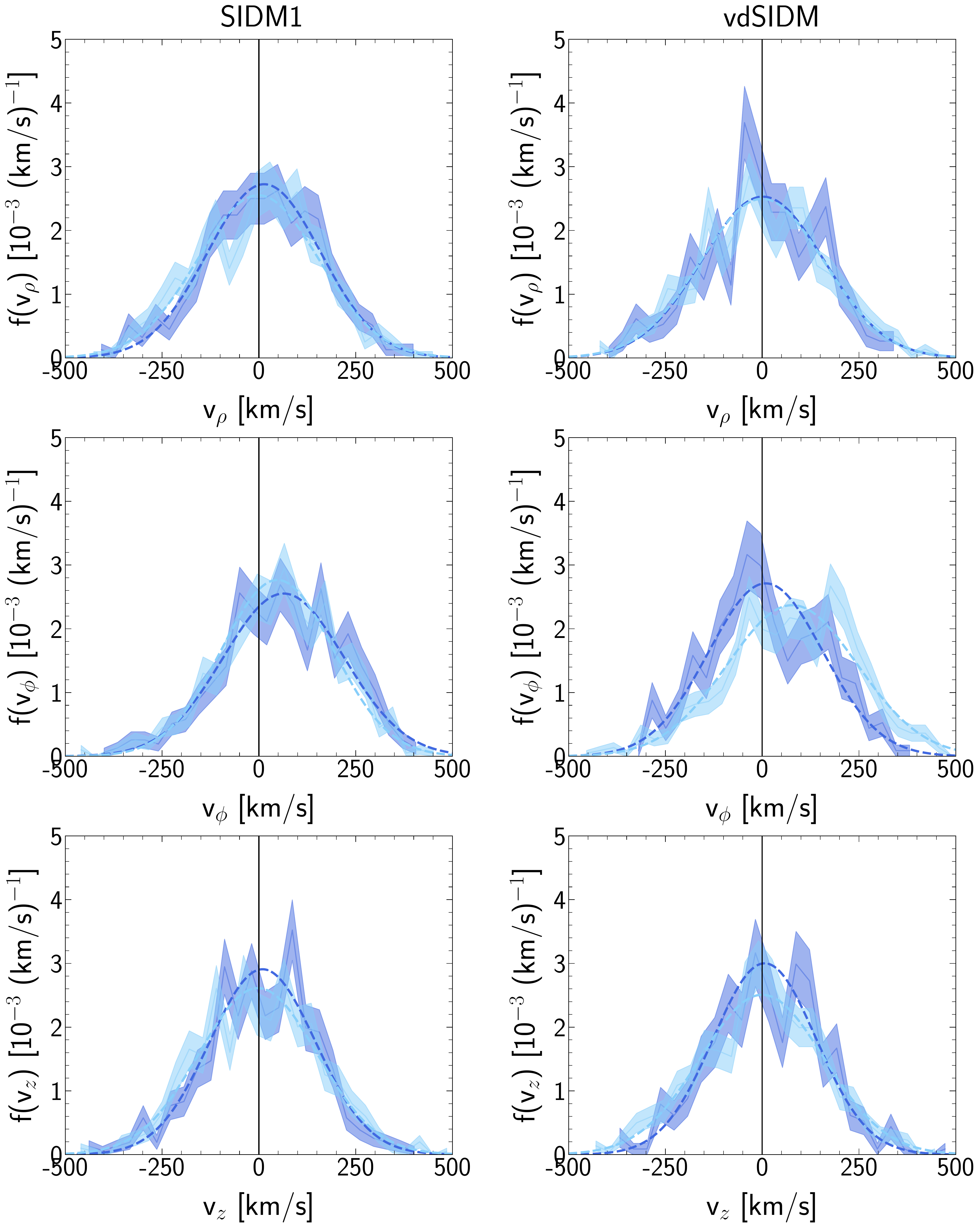}
  \caption{The radial (upper panels), azimuthal (middle panels) and vertical (bottom panels) components of the local DM velocity distributions (solid blue curves) and their 1$\sigma$ uncertainties (shaded blue bands) for the SIDM1 and vdSIDM halos that have the best (dark blue) and worst (light blue) fit velocity modulus distributions to the truncated Maxwellian distribution. The coloured dashed lines correspond to the best fit Gaussian distribution for each halo.}
  \label{fig:best_worst_component}
\end{figure}

Figure~\ref{fig:best_worst_component} shows the radial (upper panels), azimuthal (middle panels) and vertical (bottom panels) components of the local DM velocity distribution (solid blue curves)  and their 1$\sigma$ uncertainties (shaded blue bands) for the same SIDM1 and vdSIDM halos whose velocity modulus distributions are shown in figure~\ref{fig: best_worst_vel_dist}, and have the best (dark blue) and worst (light blue) fits to the truncated Maxwellian distribution. 
The best fit Gaussian distributions are shown as coloured dashed lines in each panel. As it can be seen from  the figure, the three components of the velocity distribution of the SIDM halos are not identical, and some velocity anisotropy is evident in the Solar neighbourhood of these halos. In particular, the mean azimuthal speeds of the halos are nonzero and shifted to positive values. This may be due to baryonic processes or the existence of merger events causing an offset from a mean azimuthal velocity of zero.

\begin{figure}[t]
\centering
  \begin{subfigure}[b]{1\textwidth}
  \includegraphics[width=\textwidth]{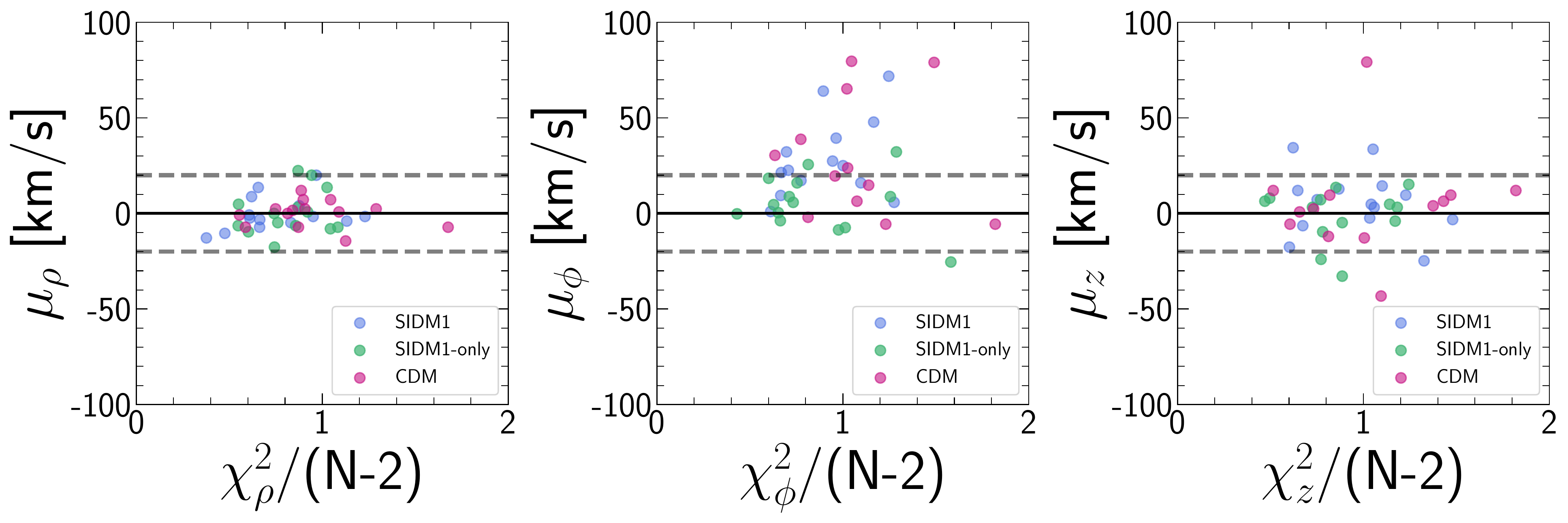}
  \end{subfigure}
    \hfill
  \begin{subfigure}[b]{1\textwidth}
  \includegraphics[width=\textwidth]{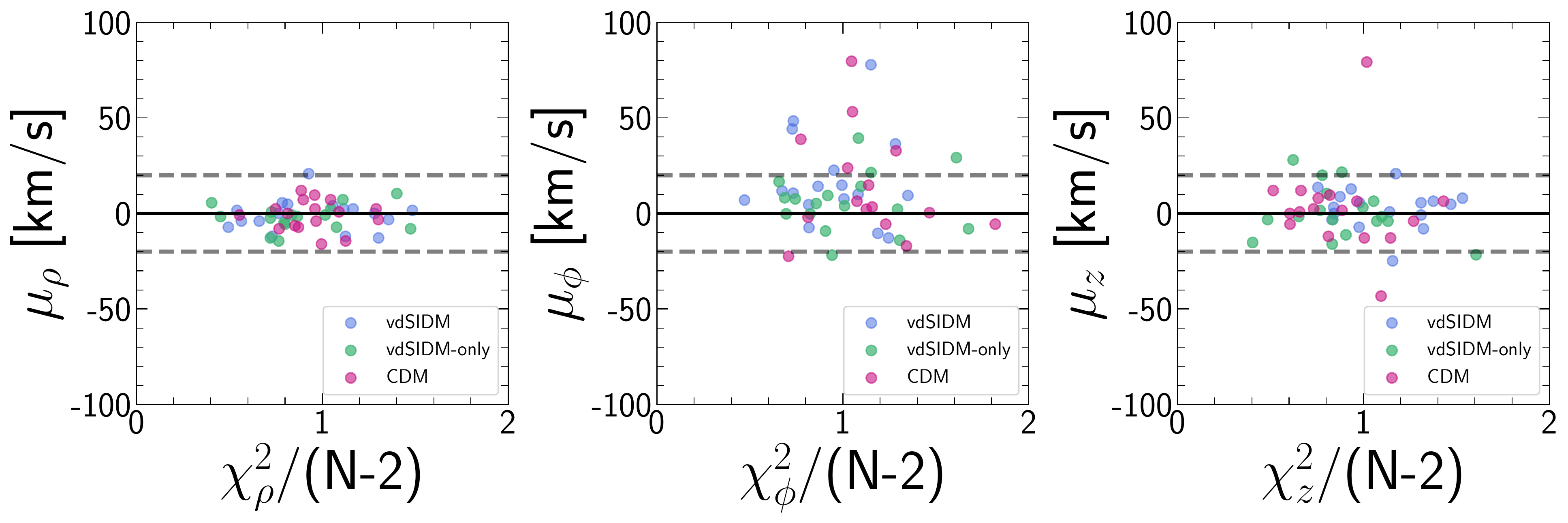}
  \end{subfigure}
  \caption{Best fit Gaussian mean velocities for the radial (left panels), azimuthal (middle panels) and vertical (right panels) components of the velocity distributions and the reduced $\chi^2$ values of the Gaussian fits for the SIDM1 (top row) and vdSIDM (bottom row) halos (blue) and their corresponding SIDM-only (green) and CDM (magenta) counterparts. 
}
  \label{fig:component_params}
\end{figure}

To further explore the velocity anisotropies and the goodness-of-fit to the Gaussian distributions, in figure~\ref{fig:component_params} we show the best fit mean velocity for each component of the velocity distribution versus the reduced $\chi^2$ values for the fits to the Gaussian distribution for halos in the EAGLE-50. The top row shows the results for the SIDM1 halos along with their SIDM1-only and CDM counterparts, while the bottom row shows the results for the vdSIDM halos and their vdSIDM-only and CDM counterparts. The left, middle, and right panels show these parameters for the radial, azimuthal and vertical velocity distribution components, respectively. As a reference, the solid black line indicates the zero mean velocity and the dashed black lines show the +20 and $-20$~km/s mean velocities in each case. The best-fit parameters for the Gaussian fits to the radial, azimuthal, and vertical velocity components for each halo are given in tables~\ref{tab:eagle-50-comp-sidm1}--\ref{tab:eagle-50-comp-vdsidm-CDM} of appendix~\ref{app:parameters}

For the SIDM1 and vdSIDM halos and their CDM counterparts, the magnitudes of the best fit Gaussian mean DM velocities in the radial direction are smaller than 20~km/s and largely concentrated around zero, whereas in the vertical direction they are slightly more spread out around the zero point.
For the azimuthal component, there are noticeably more halos with large positive best fit Gaussian mean azimuthal velocity (in the direction of the rotation of the stellar disk), with the highest velocities ($\geq$ 20 km/s) being found for halos in the hydrodynamic simulations. 
Notice that the Galactic reference frame of the SIDM-only halos are matched to their SIDM counterparts. To check if the SIDM-only halos show an azimuthal rotation about a different axis, we also consider an an alternative Galactic frame, with the $z$-axis in the direction of the average angular momentum of the DM particles. In this alternative frame, we still find the same generally lower mean azimuthal DM velocities for the SIDM-only halos. This agrees with the fact that SIDM-only simulations generally have slower moving particles due to a shallower gravitational potential well, and also suggests that baryonic processes are responsible for the rotation of the local DM distribution in the CDM and SIDM halos in the azimuthal direction.

\section{Halo integrals}\label{halo}

In this section we investigate the implications of DM self-interactions for the interpretation of DM direct detection signals. Direct detection experiments search for the small nuclear recoil energy, $E_R$, of a target nucleus in an underground detector due to scattering with a DM particle $\chi$. The differential event rate (per unit energy, per unit detector mass, per unit time) is defined as 
\begin{equation}\label{eq eventrate}
   \frac{dR}{dE_R} = \frac{\rho_\chi}{m_\chi} \frac{1}{m_A}\int_{v>v_{\rm min}} \frac{d\sigma_A}{dE_R}\,v\,\tilde{f}_{\rm det}({\bf v},t)\;d^3v
\end{equation}
where $\rho_\chi$ is the local DM density, $m_\chi$ is the DM mass, $m_A$ is the target nucleus mass, $\sigma_A$ is the DM-nucleus scattering cross section, ${\bf v}$ is the relative velocity between DM and the nucleus (with $v=|{\bf v}|$), and $\tilde{f}_{\rm det}({\bf v},t)$ is the local DM velocity distribution in the Earth's reference frame. For elastic DM-nucleus scattering, the minimum speed, $v_{\rm min}$, required for a DM particle to deposit a recoil energy $E_R$ in the detector is given by
\begin{equation}
v_{\rm min} = \sqrt{\frac{m_A E_R}{2 \mu_{\chi A}^2}},
\label{eq:earth detector}
\end{equation}
where $\mu_{\chi A} = m_{\chi} m_A / (m_{\chi} + m_A)$ is the DM-nucleus reduced mass. 

The astrophysical dependence of the event rate is encoded in the local DM density, $\rho_\chi$, and the local DM velocity distribution in the detector frame, $\tilde{f}_{\rm det}({\bf v}, t)$ through an integral over the DM velocity. For standard spin-independent and spin-dependent DM-nucleus interactions, the event rate is proportional to the so-called \emph{halo integral},
\begin{equation}
    \eta(v_{\textrm{min}}, t) = \int_{v>v_{\textrm{min}}} d^3v \hspace{1mm} \frac{\tilde{f}_{\textrm{det}}  (\textbf{v}, t)}{v}.
\end{equation}
Thus, studying the halo integral is  crucial to understand and quantify the astrophysical dependence of the results of direct detection experiments and explore how DM self-interactions would affect their interpretation. 

To compute the halo integrals for the simulated halos, we first need to transform the local DM velocity distribution from the Galactic frame to the detector frame such that,
\begin{equation}
\tilde{f}_{\textrm{det}}(\textbf{v},t) = \tilde{f}_{\textrm{gal}}(\textbf{v}+\textbf{v}_{s}+\textbf{v}_e(t)),    
\end{equation}
where $\textbf{v}_e(t)$ is the velocity of the Earth with respect to the Sun and $\textbf{v}_{s}$ is the Sun's velocity with respect to the Galactic center. The latter is given by $\textbf{v}_s = \textbf{v}_c + \textbf{v}_{\textrm{pec}}$, where $\textbf{v}_c$ is the Sun's circular velocity and $\textbf{v}_{\textrm{pec}} \approx (11.10,12.24,7.25)$ km/s \cite{Schoenrich:2009bx} is the peculiar velocity of the Sun with respect to the Local Standard of Rest. The velocity of the Earth with respect to the Sun, $\textbf{v}_e(t)$, introduces a time dependence in the halo integral. In the following, we compute the  time-averaged  halo integrals, which are averaged over one year.

\begin{figure}[t]
 \centering
    \includegraphics[width=0.8\textwidth]{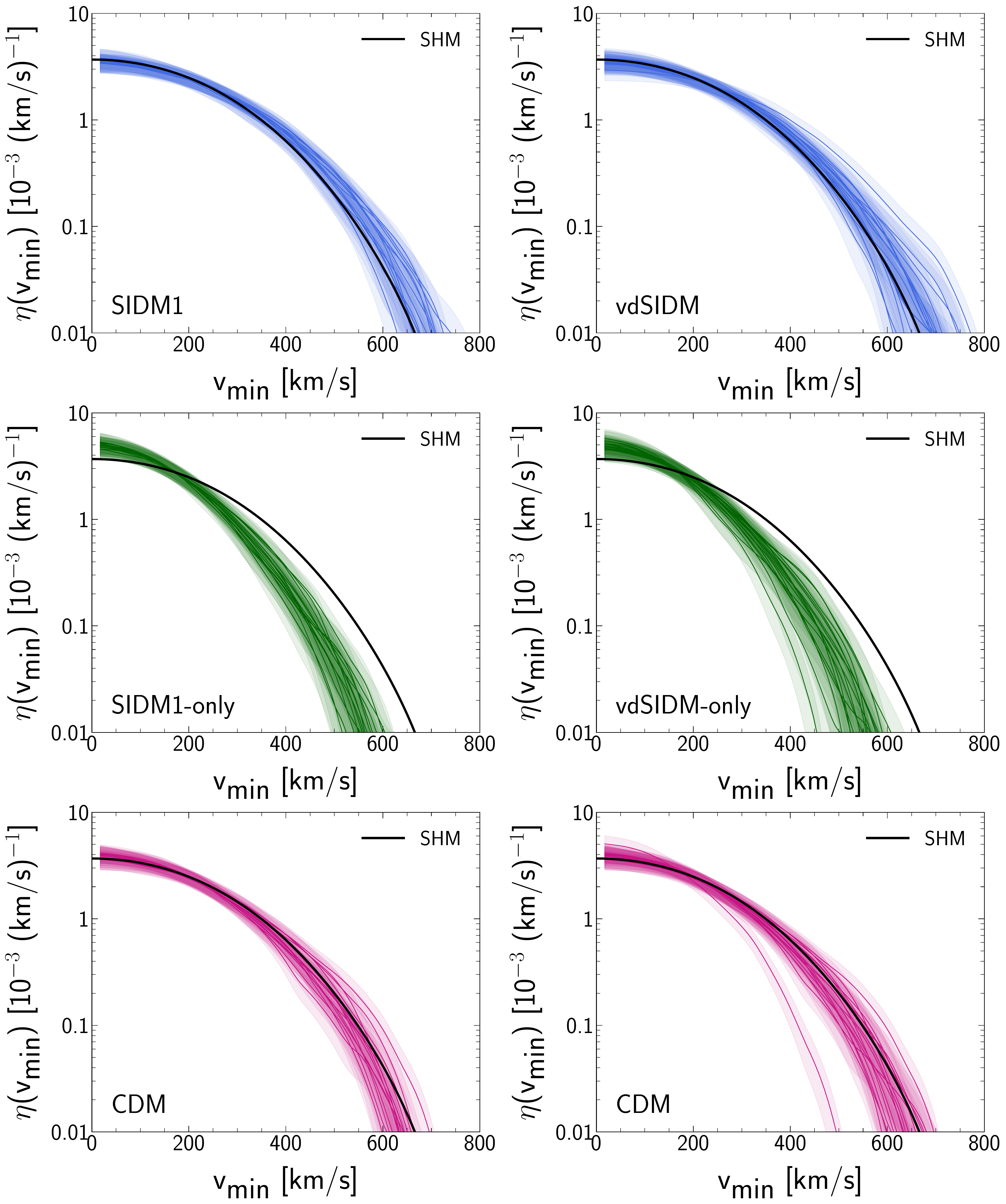}
  \caption{Time-averaged halo integrals (solid coloured lines) and their 1$\sigma$ uncertainties (coloured bands) as a function of the DM minimum speed, $v_{\rm min}$, for all the selected MW-like SIDM1 (top left panel)  and vdSIDM (top right panel) halos, along with their SIDM-only (middle panels) and CDM (bottom panels) counterparts in the EAGLE-50 simulations. The solid black curves show the halo integral obtained from the SHM Maxwellian velocity distribution with a peak speed of 230~km/s. 
}
  \label{all_halo_int}
\end{figure}

We next proceed to extract the halo integrals of the simulated halos. In figure \ref{all_halo_int} we present the time-averaged halo integrals (solid coloured lines) and their 1$\sigma$ uncertainties (shaded bands) for all selected MW-like SIDM1 and vdSIDM halos (top panels) in  EAGLE-50. The solid coloured lines are obtained by computing the halo integrals from the mean value of the local DM velocity distributions, while the uncertainty bands are determined by adding and subtracting $1\sigma$ Poisson uncertainty to the mean velocity
distributions. The corresponding SIDM-only and CDM counterparts are shown in the middle and bottom panels, respectively. The black solid curve shows the halo integral for the SHM Maxwellian velocity distribution with a peak speed of 230~km/s. 

The inclusion of baryons in the simulations results in the tails of the halo integrals shifting to higher $v_{\rm min}$ values, which is due to the shifts observed in the local DM velocity distributions (seen in figure~\ref{all_vel_dist}). Including  DM self-interactions along with baryons further shifts the tails to even higher DM minimum speeds. This is due to the slight shift of the DM velocity distributions of SIDM halos to higher speeds compared to their CDM counterparts, which are not all MW-like. 
In figure~\ref{fig: cdm_speed_and_halo} of appendix~\ref{app:CDM}, for comparison we present the halo integrals of the CDM halos that satisfy our MW-like halo selection criteria. We also note that there are no significant differences between the SIDM1 and vdSIDM halo integrals. The only difference observed is that the halo-to-halo scatter is larger in the vdSIDM halo integrals most probably because they are more numerous compared to the SIDM1 halos. 

Next, we study how well the halo integral obtained from the truncated Maxwellian distribution fits the halo integrals extracted from the simulations. The analytic halo integral of the truncated Maxwellian velocity distribution is given by \cite{Savage:2006qr}.
\begin{equation}
\eta(v_{\rm min},t) =\begin{cases}
\frac{1}{v_0 y}   , & {\rm for}~ z<y,~x<|y-z| \\
\frac{1}{2N_{\rm esc} v_0 y}\left[\erf(x+y)-\erf(x-y)-\frac{4}{\sqrt{\pi}}y e^{-z^2}\right], & {\rm for}~z>y,~x<|y-z|\\
\frac{1}{2N_{\rm esc} v_0 y}\left[\erf(z)-\erf(x-y)-\frac{2}{\sqrt{\pi}}(y+z-x) e^{-z^2}\right], & {\rm for}~|y-z|<x<y+z\\
0, & {\rm for}~y+z<x
\end{cases}
\end{equation}
where $x = v_{\rm min}/v_0$, $ y = v_E(t)/v_0$, $z = v_{\rm esc}/v_0$, $v_0$ is the peak speed of the speed distribution, and $v_E(t)=|{\bf v}_s + {\bf v}_e(t)|$.

\begin{figure}[t]
 \centering
    \includegraphics[width=0.8\textwidth]{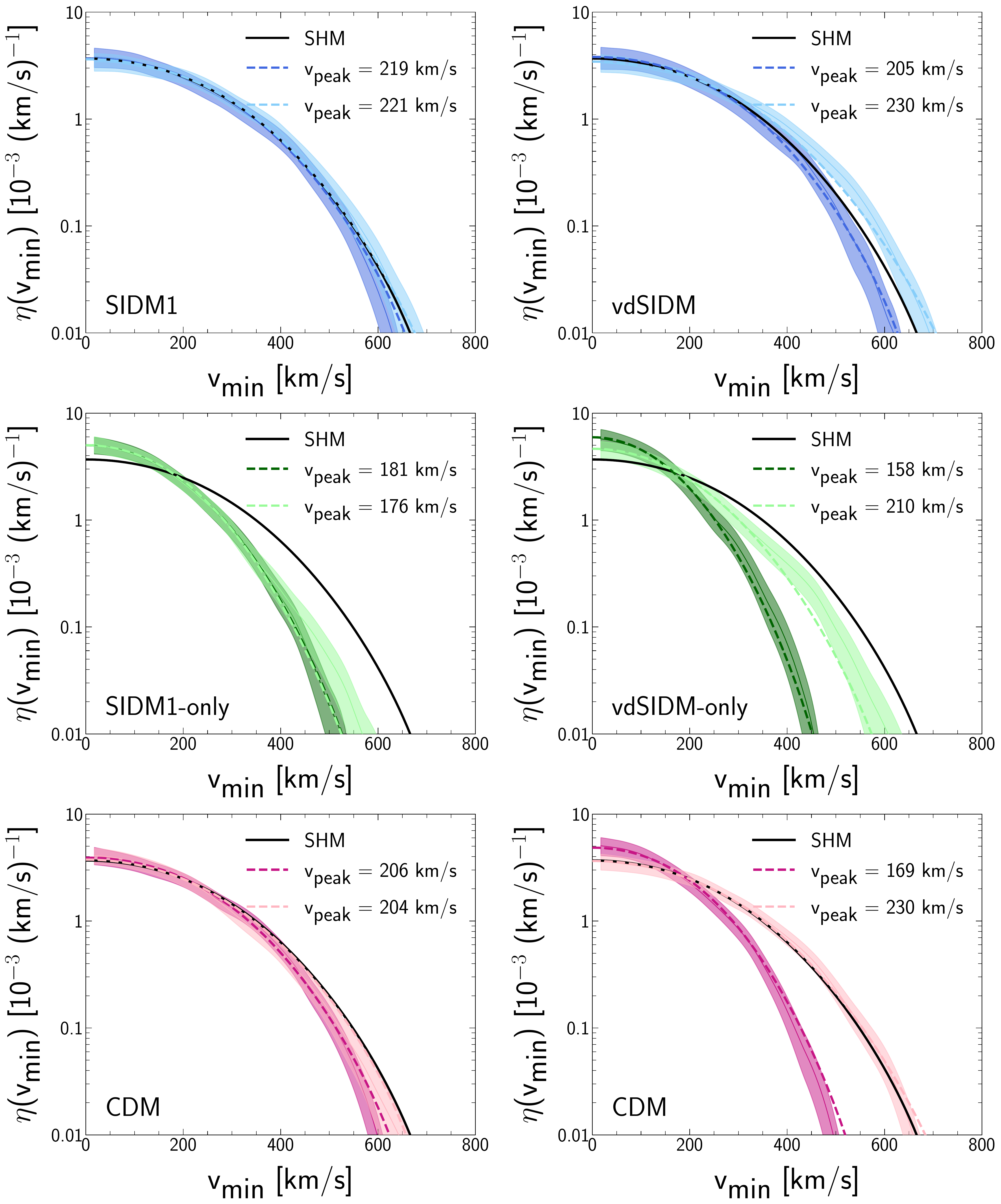}
  \caption{Time averaged halo integrals (solid coloured lines) and their 1$\sigma$ uncertainties (shaded bands) for the same two SIDM1 (top left panel) and  vdSIDM (top right panel) halos whose DM velocity modulus distributions are shown in  figure~\ref{fig: best_worst_vel_dist} and have the best (dark blue) and worst (light blue) fit to the truncated Maxwellian distribution. The halo integrals for the corresponding SIDM-only (green) and CDM (magenta) halos are shown in the middle and bottom panels, respectively. The coloured dashed lines show the best fit Maxwellian halo integrals for each halo, with the best fit peak speeds of the corresponding Maxwellian speed distributions indicated on each panel. The black solid line corresponds to the SHM halo integral with a peak speed of 230~km/s.}
  \label{fig:6.2}
\end{figure}

For each halo, we compute the analytic halo integral using the best fit peak speed of its corresponding Maxwellian velocity modulus distribution. 
Figure~\ref{fig:6.2} shows the time-averaged halo integrals (solid coloured lines) and their 1$\sigma$ uncertainties (shaded bands) for the same two SIDM1 and vdSIDM halos (top panels) whose velocity modulus distributions are shown in figure~\ref{fig: best_worst_vel_dist}, and have the best (dark colour)  and worst (light colour)  fits to the truncated Maxwellian distribution. The corresponding SIDM-only and CDM counterparts are shown in the middle and bottom panels, respectively.  
The dashed coloured lines in each panel specify the best fit Maxwellian halo integral for each halo, with the best fit peak speeds of the corresponding Maxwellian speed distributions also quoted. The solid black line shows the halo integral of the SHM Maxwellian velocity distribution with a peak speed of 230~km/s. As can be seen from the figure, the best fit Maxwellian halo integrals lie within the 1$\sigma$ uncertainty band of the halo integrals extracted from the simulations over the full $v_{\rm min}$ range for both the best and worst fit SIDM1 and vdSIDM halos, along with their corresponding SIDM-only and CDM halos.

We also investigate how well the best fit Maxwellian halo integrals fit those extracted from the simulations for all our selected MW-like halos.
From the 14 MW-like SIDM1 halos in the EAGLE-50 simulations, the best fit Maxwellian halo integrals lie within the 1$\sigma$ uncertainty bands over the full $v_{\rm min}$ range for 9 SIDM1 halos, as well as 11 SIDM1-only and 8 CDM counterparts. The remaining SIDM1, SIDM1-only, and CDM halos have best fit Maxwellian halo integrals which begin deviating from the 1$\sigma$ uncertainty band at $v_{\rm min} \sim [575 - 650]$~km/s, $[500 - 600]$~km/s, and $[575 - 650]$~km/s, respectively. When considering all 17 vdSIDM halos in EAGLE-50, the best fit Maxwellian halo integrals lie within the 1$\sigma$ uncertainty bands over the full $v_{\rm min}$ range for 13 vdSIDM halos, as well as 15 vdSIDM-only and 12 CDM counterparts. The remaining vdSIDM, vdSIDM-only, and CDM halos  have best fit Maxwellian halo integrals which begin deviating from the 1$\sigma$ uncertainty band at $v_{\rm min} \sim [600 - 650]$~km/s, $[500 - 525]$~km/s, and $[575 - 650]$~km/s, respectively.

The range of the halo-to-halo scatter in the halo integrals translates to astrophysical uncertainties in the exclusion limits set by direct detection experiments in the plane of DM mass and cross section~\cite{Bozorgnia:2016ogo, Bozorgnia:2017brl, Bozorgnia:2019mjk}. In particular, uncertainties in the halo integrals at large $v_{\rm min}$ introduce uncertainties in the direct detection exclusion limits at low DM masses, where the experiments probe the high speed tail of the halo integrals. As it can be seen from figures~\ref{all_halo_int} and \ref{fig:6.2}, the halo integrals of SIDM halos are slightly shifted to larger speeds compared to the CDM halos, which leads to a shift of the direct detection exclusion limits towards smaller DM masses.

\section{Conclusions}\label{conclusions}

In this work, we have studied the local DM density and velocity distribution of a set of MW-like galaxies in SIDM versions~\cite{Robertson:2020pxj, Robertson:2017mgj} of the EAGLE simulations~\cite{Schaye:2015, Crain:2015poa}, and explored their implications for DM direct detection. We considered two SIDM models; DM with a velocity-independent and isotropic self-interacting cross section of 1 cm$^2$~g$^{-1}$ (SIDM1), and DM with a velocity-dependent and anisotropic cross section corresponding to scattering through a Yukawa potential (vdSIDM). For the two SIDM simulations, CDM (collisionless DM + baryons) and SIDM-only (self-interacting DM only) counterparts of the  halos were also available, and we compare and contrast our results among all these cases.

We first identified MW-like simulated halos 
according to four criteria: (i) the virial mass of the simulated galaxy  is within $[0.5 - 3.0] \times 10^{12}~\Msun$~\cite{Callingham:2018vcf}, (ii) its stellar mass is within the 3$\sigma$ range of the observed MW stellar mass, $[4.5 - 8.3] \times 10^{10}~\Msun$~\cite{McMillan:2011wd}, (iii) its rotation curve fits well the observed MW rotation curve~\cite{Iocco:2015xga}, and (iv) the simulated halo is relaxed. 
Our criteria yielded 14 SIDM1 and 17 vdSIDM halos. 
Our main findings are summarized below:

\begin{itemize}

  \item \textbf{Local DM density:} The local DM densities of the MW-like SIDM1 and vdSIDM halos are in the range of $[0.41-0.66]$ and $[0.30-0.67]$~GeV/cm$^3$, respectively. 
  The CDM halo have comparable local DM densities to their SIDM counterparts, while some of the SIDM-only halos have slightly smaller local DM densities.

  \item \textbf{Local DM velocity distributions:} The local DM velocity modulus distributions of the SIDM1 and vdSIDM halos are similar, and generally agree with those extracted from their CDM counterparts. They exhibit only slight shifts to higher speeds compared to their CDM counterparts, due to the lower rotation curves of the CDM halos and the fact that they are not all MW-like. The velocity modulus distributions of the SIDM-only halos, on the other hand, have generally lower peak speeds than both the SIDM and CDM halos, indicating the prominent effect of baryons in increasing the DM speeds.   The  truncated Maxwellian speed distribution provides a good fit to the local DM speed distributions of  the SIDM1, vdSIDM, CDM, SIDM1-only, and vdSIDM-only halos. 
  Moreover, the speed distributions of the SIDM1, vdSIDM and CDM halos are in general agreement with the SHM Maxwellian speed distribution.

 \item \textbf{Halo integrals:} The halo integrals obtained from the best fit truncated Maxwellian velocity distribution for the SIDM1, vdSIDM, CDM, SIDM1-only, and vdSIDM-only  halos provide a good fit to the halo integral obtained directly from the simulations, without any systematic difference between the results of different models. Small deviations from the simulated halo integral only begin occurring at relatively high values of $v_{\rm min}$. The large halo-to-halo variation in the halo integrals at large $v_{\rm min}$ leads to large astrophysical uncertainties in direct detection exclusion limits at low DM masses.

\end{itemize}

Our results show that including DM self-interactions in hydrodynamic simulations does not in general  lead to any significant difference between the local DM densities, local DM speed distributions, or halo integrals compared to collisionless CDM simulations including baryons. In both cases, a Maxwellian velocity distribution with a free peak speed fits well the local DM speed distributions of the halos. We have also shown that including baryons (with or without DM self-interactions) in the simulations has a major effect on the local DM distribution. This agrees with the results of previous work, which compared the local DM distributions in CDM (including baryons) and DM-only halos in various high resolution cosmological simulations~\cite{Poole-McKenzie:2020dbo,  Bozorgnia:2016ogo, Bozorgnia:2017brl, Bozorgnia:2019mjk, Kelso:2016qqj, Sloane:2016kyi}.

\subsection*{Acknowledgements}
We thank Lukas Arda, Fabian Bautista, and Jasmin Hartmann, who participated in the initial phases of this research project through the EXPLORE program. ER and NB acknowledge the support of the Natural Sciences and Engineering Research Council of Canada (NSERC), funding reference number RGPIN-2020-07138, the NSERC Discovery Launch Supplement, DGECR-2020-00231, and the Academic Innovation Fund at York University. EV and NB acknowledge the support from the HQP Pooled Resources from the McDonald Institute, application number HQP 2020-R6-01. This work used the DiRAC Memory Intensive system at Durham University, operated by ICC on behalf of the STFC DiRAC HPC Facility (www.dirac.ac.uk). This equipment was funded by BIS National E-infrastructure capital grant ST/K00042X/1, STFC capital grant ST/H008519/1, and STFC DiRAC Operations grant ST/K003267/1 and Durham University. DiRAC is part of the National E-Infrastructure.


\clearpage

\section*{Appendix}
\addcontentsline{toc}{section}{Appendix}
\renewcommand{\thesubsection}{\Alph{subsection}}
\renewcommand{\theequation}{\thesubsection.\arabic{equation}}

\subsection{Rotation curve goodness of fit}\label{App.RC}

In this appendix we present our analysis for fitting the rotation curves of the simulated halos to the observed MW rotation curve. 
We follow the procedure given in ref.~\cite{Calore:2015oya} to determine the goodness of fit, and minimize the two-variable $\chi^2$ given by~\cite{Iocco:2015xga}
\begin{equation}
\chi^2 = \sum_{i} \frac{(y_i - \hat{y}(x_i))^2}{\sigma_{y_i}^2 + (dy_i/dx_i)^2\sigma_{x_i}^2},
\end{equation}
where $i$ runs over the observational data points considered, $x \equiv r/R_0$, $y \equiv \omega_c/\omega_0 -1$, and $\omega_0 \equiv v_0/R_0$. We adopt a local galactocentric distance of $R_0=8$~km/s and a local circular speed of $v_0=230$~km/s. Here, $\hat{y}(x_i)$ is the simulated rotation curve evaluated at $x_i$, and $\sigma_{x_i}$ and $\sigma_{y_i}$ are the 1$\sigma$  errors in the observational data.

One of the criteria to select a simulated halo as MW-like  is for its rotation curve to broadly agree with the observed MW-rotation curve. We therefore select only halos which have a reduced $\chi^2$, $\chi^2/(N-1) \leq 200$, where $N=2687$ is the total number of observational data points. The reduced $\chi^2$ values for the selected MW-like SIDM and vdSIDM halos in the EAGLE-50 simulations is given in table~\ref{table of mw reduced chi squared}.

\begin{table}[h!]
\centering
\resizebox{0.32\columnwidth}{!}{%
\renewcommand{\arraystretch}{1.19}

\begin{tabular}[]{ |c|c|  }
 \hline
 \multicolumn{2}{|c|}{SIDM1} \\
 \hline
 Halo Number & $\chi^2/(N-1)$ \\
 \hline
 48 & 157  \\
 60 & 32.8 \\
 71 & 19.1  \\
 74 & 40.8  \\
 80 & 38.3 \\
 92 & 54.5  \\
 96 & 65.4  \\
 99 & 20.3  \\
 102 & 16.2  \\
 111 & 48.5  \\
 118 & 22.3 \\
 128 & 68.9  \\
 131 & 61.1  \\
 159 & 19.8  \\
 \hline
\end{tabular}%
}
\resizebox{0.32\columnwidth}{!}{%
\renewcommand{\arraystretch}{1}
\begin{tabular}[]{ |c|c|  }
 \hline
 \multicolumn{2}{|c|}{vdSIDM} \\
 \hline
 Halo Number &$\chi^2/(N-1)$\\
 \hline
 55 & 126  \\
 60 & 34.3  \\
 72 & 143  \\
 74 & 51.2  \\
 75 & 101  \\
 77 & 95.6  \\
 92 & 163  \\
 93 & 32.8  \\
 99 & 115  \\
 101 & 40.8  \\
 107 & 49.8  \\
 115 & 58.4 \\
 122 & 130 \\
 124 & 35.9 \\
 128 & 133  \\
 138 & 129  \\
 176 & 125  \\
 \hline
\end{tabular}%
}
\captionof{table}{\label{table2} Reduced $\chi^2$ values for the   SIDM1 (left table) and vdSIDM (right table) halos satisfying our MW-like halo selection criteria.}
\label{table of mw reduced chi squared}
\end{table}

\begin{figure}[t]
  \begin{subfigure}[b]{0.5\textwidth}
    \includegraphics[width=\textwidth]{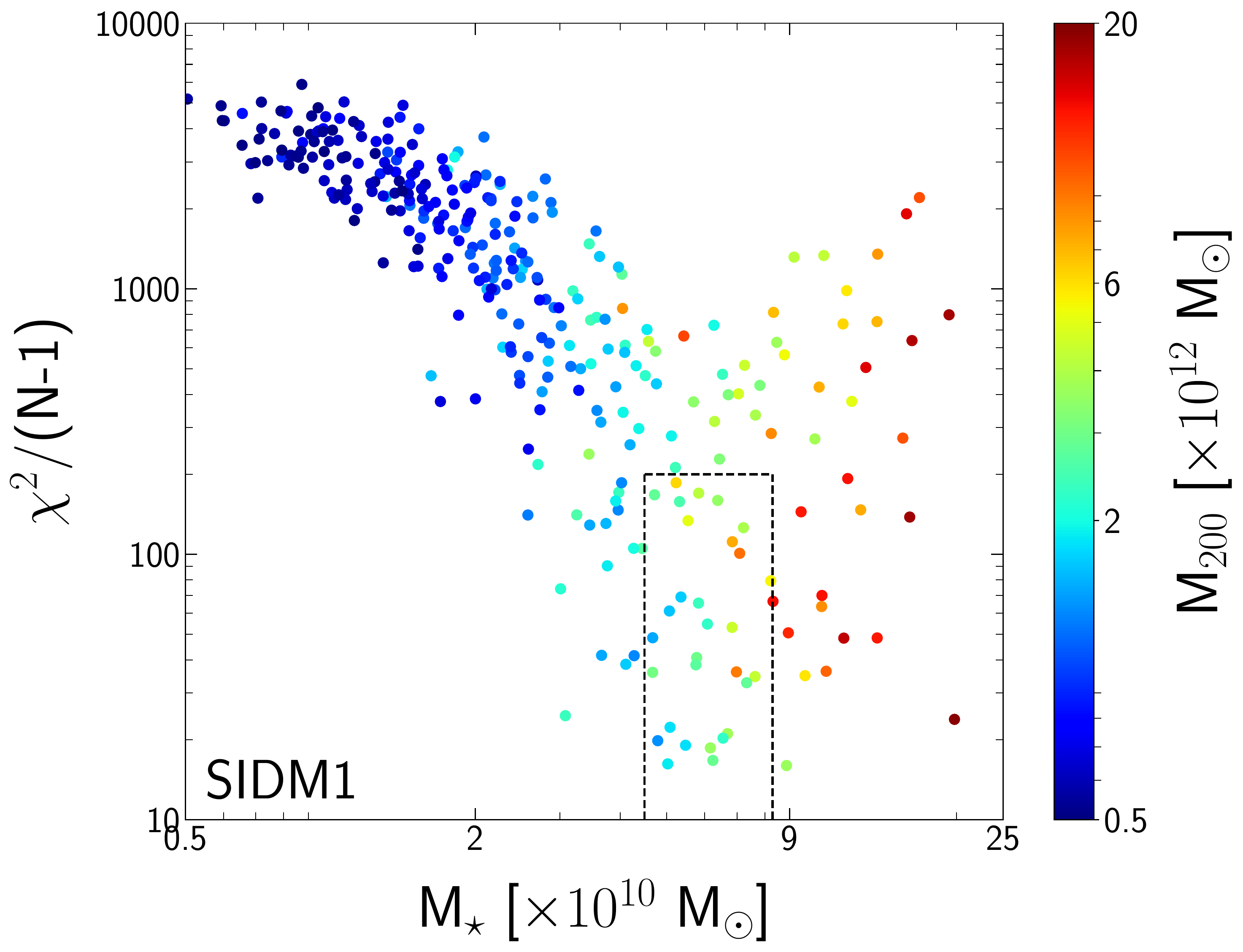}
  \end{subfigure}
  \hfill
  \begin{subfigure}[b]{0.5\textwidth}
    \includegraphics[width=\textwidth]{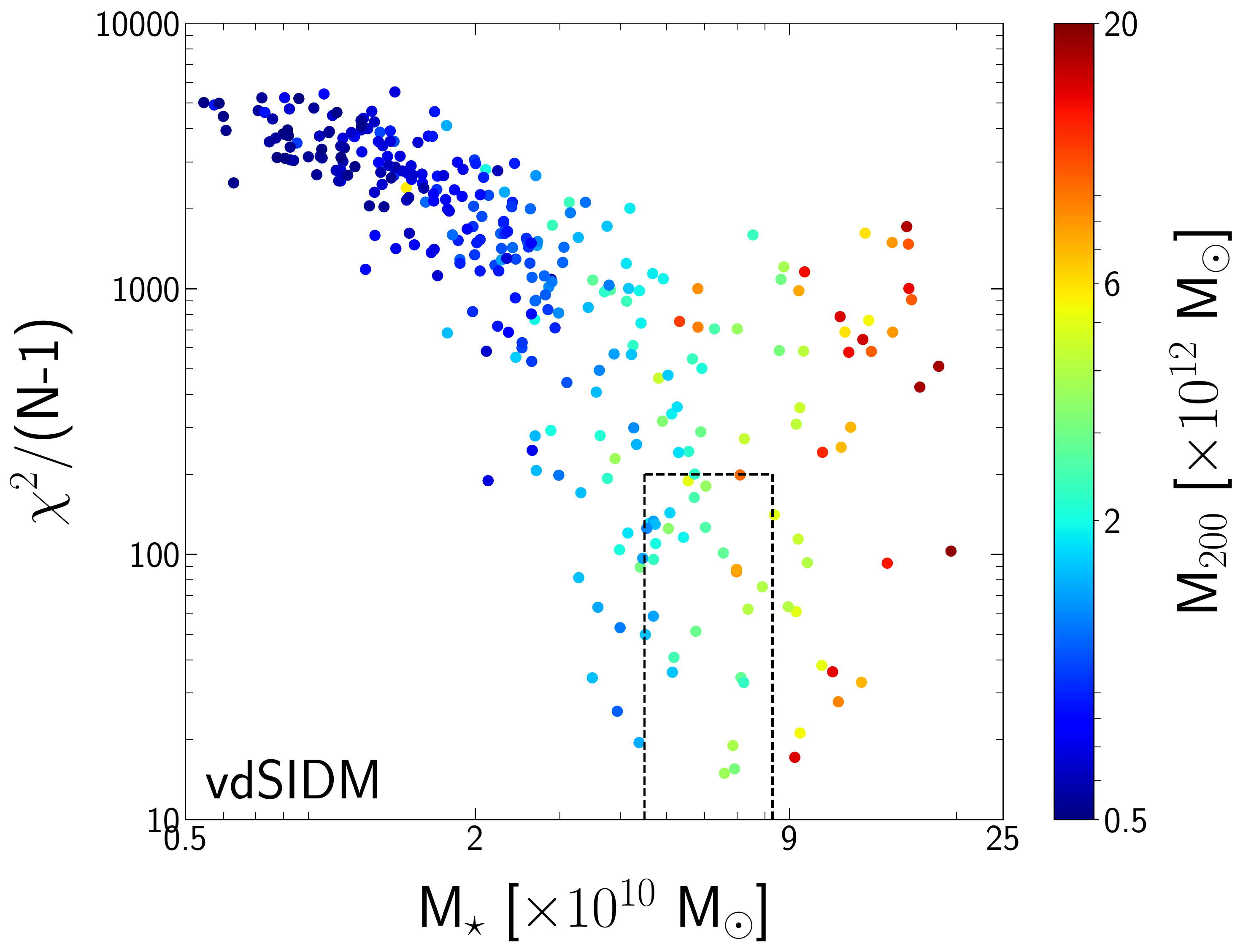}
  \end{subfigure}
  \caption{Reduced $\chi^2$ values from the rotation curve fit versus the stellar mass for the SIDM1 (left panel) and vdSIDM (right panel) halos in the EAGLE-50 simulations. The black dashed box specifies the region in which the halos have stellar masses within the 3$\sigma$ range of the observed MW stellar mass and $\chi^2/(N-1) \leq 200$. 
  The colour bar shows the distribution of the virial mass, $M_{200}$.}
  \label{fig: mstar vs chi}
\end{figure}

In figure~\ref{fig: mstar vs chi}, we show the reduced $\chi^2$ versus the stellar mass for the SIDM1 (left panel) and vdSIDM (right panel) halos. The colour bars shows the virial mass of the halos. It is clear from the figure that the global minimum of the reduced $\chi^2$ distributions occur within the 3$\sigma$ range of the observed MW stellar mass.  

\subsection{CDM Milky Way-like galaxies}\label{app:CDM}

In this appendix we investigate how the rotation curves, DM velocity modulus distributions, and halo integrals of the CDM halos change if we select only those which satisfy our MW-like galaxy selection criteria, instead of selecting the CDM counterparts of the SIDM halos.

\begin{figure}[t]
\centering
  \begin{subfigure}[b]{1\textwidth}
    \includegraphics[width=\textwidth]{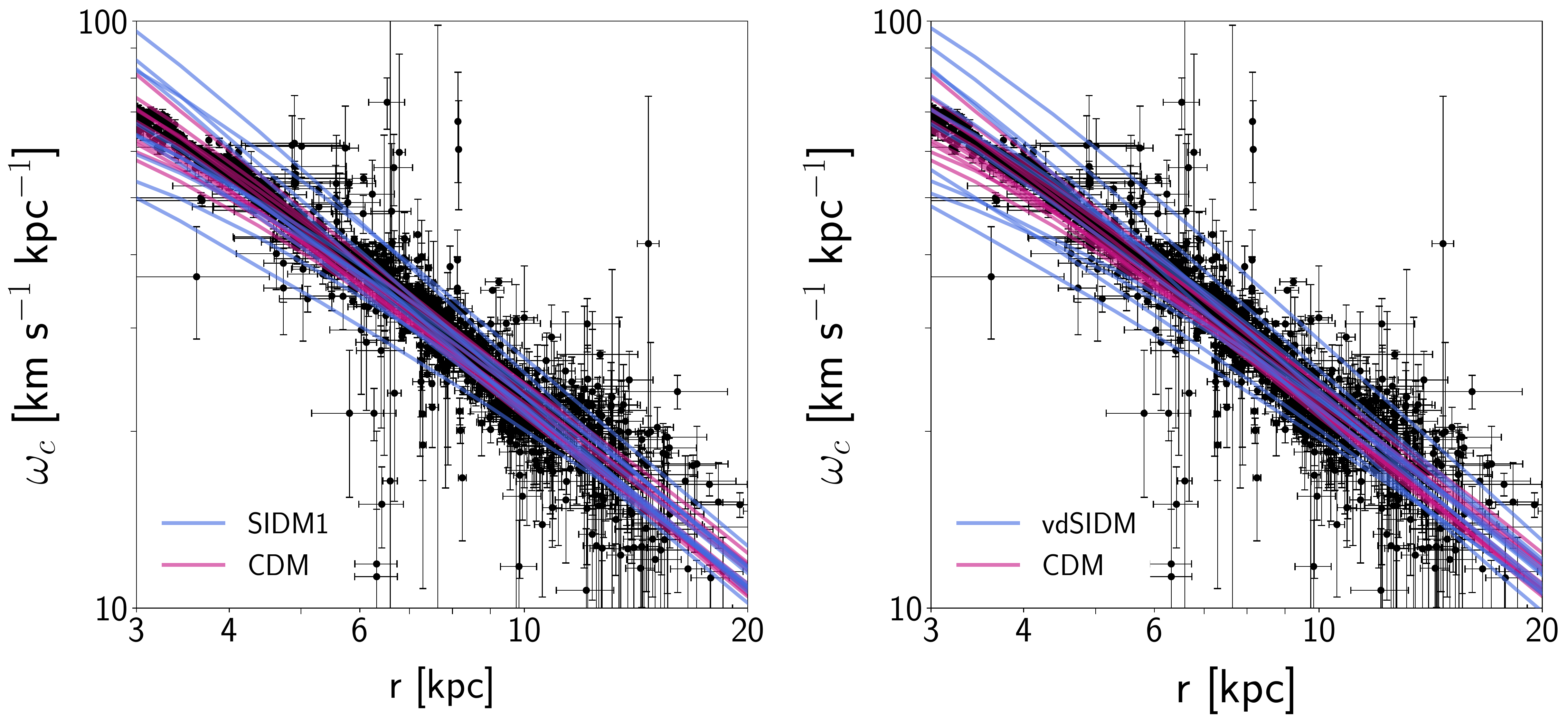}
  \end{subfigure}
  \caption{Angular circular velocity, $\omega_c$, as a function of Galactocentric distance, $r$, for the selected MW-like CDM halos (magenta curves). The rotation curves for the MW-like SIDM1 and vdSIDM halos (blue curves) are overplotted in the left  and right panels, respectively (same as shown in figure~\ref{fig:rotation curves}). The black points and their error bars represent the observational measurements of the MW rotation curve from ref.~\cite{Iocco:2015xga} and their $1\sigma$ errors. }
  \label{fig: rotation_curve_cdm}
\end{figure}

We find that 11 CDM halos in the EAGLE-50 simulations satisfy the four selection criteria outlined in section~\ref{selection}. Out of these 11 halos, only 5 (4) are counterparts of the SIDM1 (vdSIDM) halos. In figure \ref{fig: rotation_curve_cdm} we show the angular circular velocity as a function of the Galactocentric distance for the  MW-like CDM halos as magenta curves. For comparison, the rotation curves of the selected MW-like SIDM1 and vdSIDM halos are shown as blue curves in the left and right panels, respectively. These are the same as the blue curves shown in figure~\ref{fig:rotation curves}). As expected, the rotation curves of the MW-like CDM halos agree well with both the observed MW rotation curves and those obtained from the SIDM MW-like halos.

\begin{figure}[t]
\centering
  \begin{subfigure}[b]{1\textwidth}
    \includegraphics[width=\textwidth]{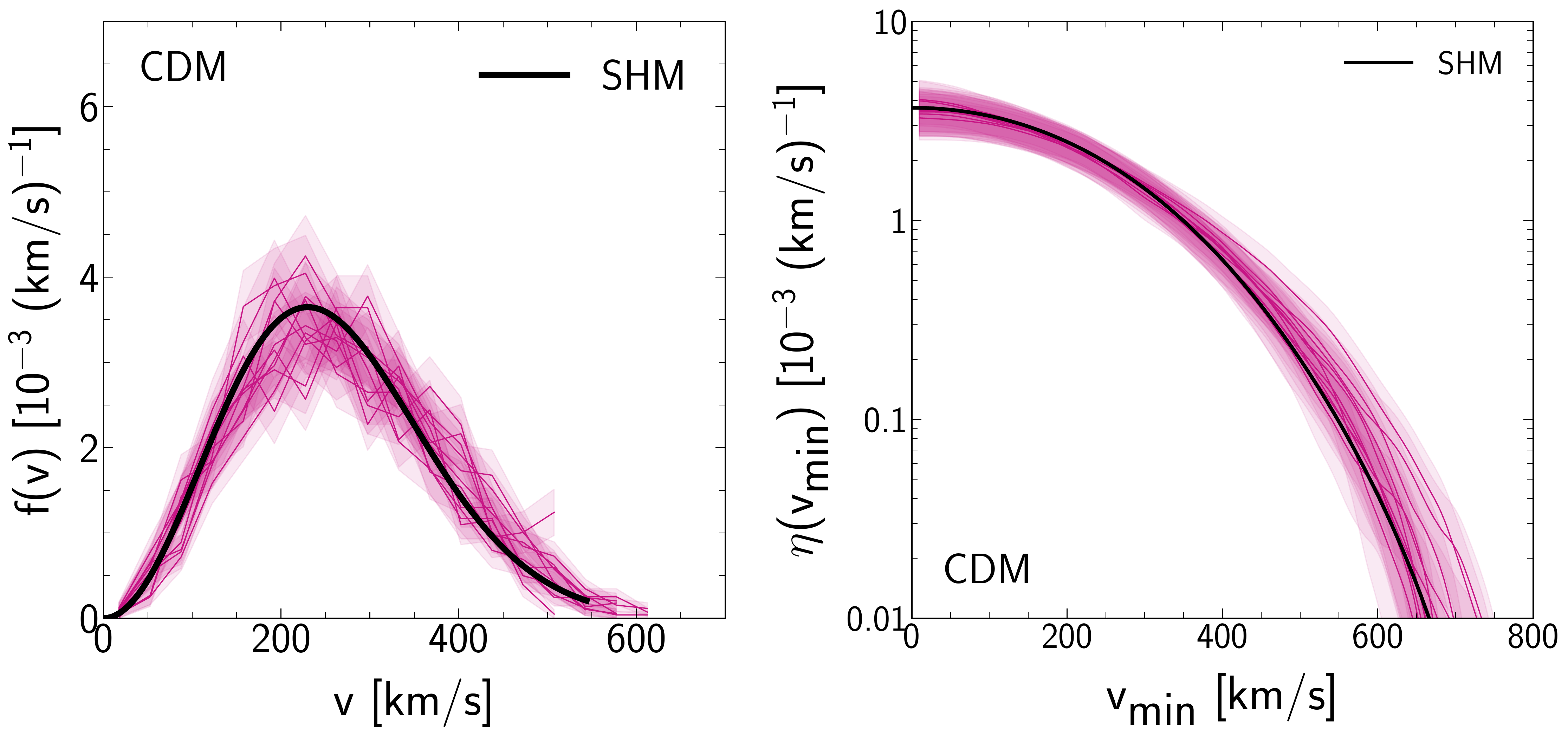}
  \end{subfigure}
  \caption{The local DM velocity modulus distributions (left panel) and the time-averaged halo integrals (right panel) with their 1$\sigma$ Poisson uncertainties for all selected MW-like CDM halos. The solid black curve specifies the SHM Maxwellian distribution with a peak speed of 230~km/s.}
  \label{fig: cdm_speed_and_halo}
\end{figure}

In the left panel of figure~\ref{fig: cdm_speed_and_halo}, we present the local DM velocity modulus distributions (solid magenta lines) and their $1\sigma$ uncertainties (shaded bands) for the 11 MW-like CDM halos. In the right panel, we present the halo integrals (solid magenta lines) and their $1\sigma$ uncertainties (shaded bands)  for the same halos. We also fit the truncated Maxwellian  distribution to the local DM velocity modulus distribution of the MW-like CDM halos, and find that the range of their best fit peak speeds is $[215 – 253]$~km/s. This is in agreement with the ranges of the peak speed of the SIDM halos presented in section~\ref{velmod}.

 \subsection{Truncated generalized Maxwellian distribution}\label{app:Gen Maxwellian}

The truncated generalized Maxwellian velocity modulus distribution is given by,
\begin{equation}
f(v) =\begin{cases}
\frac{v^2}{N_{\rm esc}^{\rm gen}} e^{-(v^2/v_0^2)^\alpha}   , & {\rm for}~v<v_{\rm esc} \\
0, & {\rm otherwise}
\end{cases}
\label{eq:genmax}
\end{equation}
\noindent where
\begin{equation}
N_{\rm esc}^{\rm gen} = v_0^3\left[\Gamma(3/2\alpha)-\Gamma(3/2\alpha,z^{2\alpha})\right]/2\alpha,  
\end{equation}
and $v_0$ and $\alpha$ are free parameters, $z=v_{\rm esc}/v_0$, and $N_{\rm esc}^{\rm gen}$ is a normalization factor such that $\int f(v) dv =1$. In the case of a standard Maxwellian distribution, $\alpha=1$.

\begin{figure}[t]
 \centering
    \includegraphics[width=0.8\textwidth]{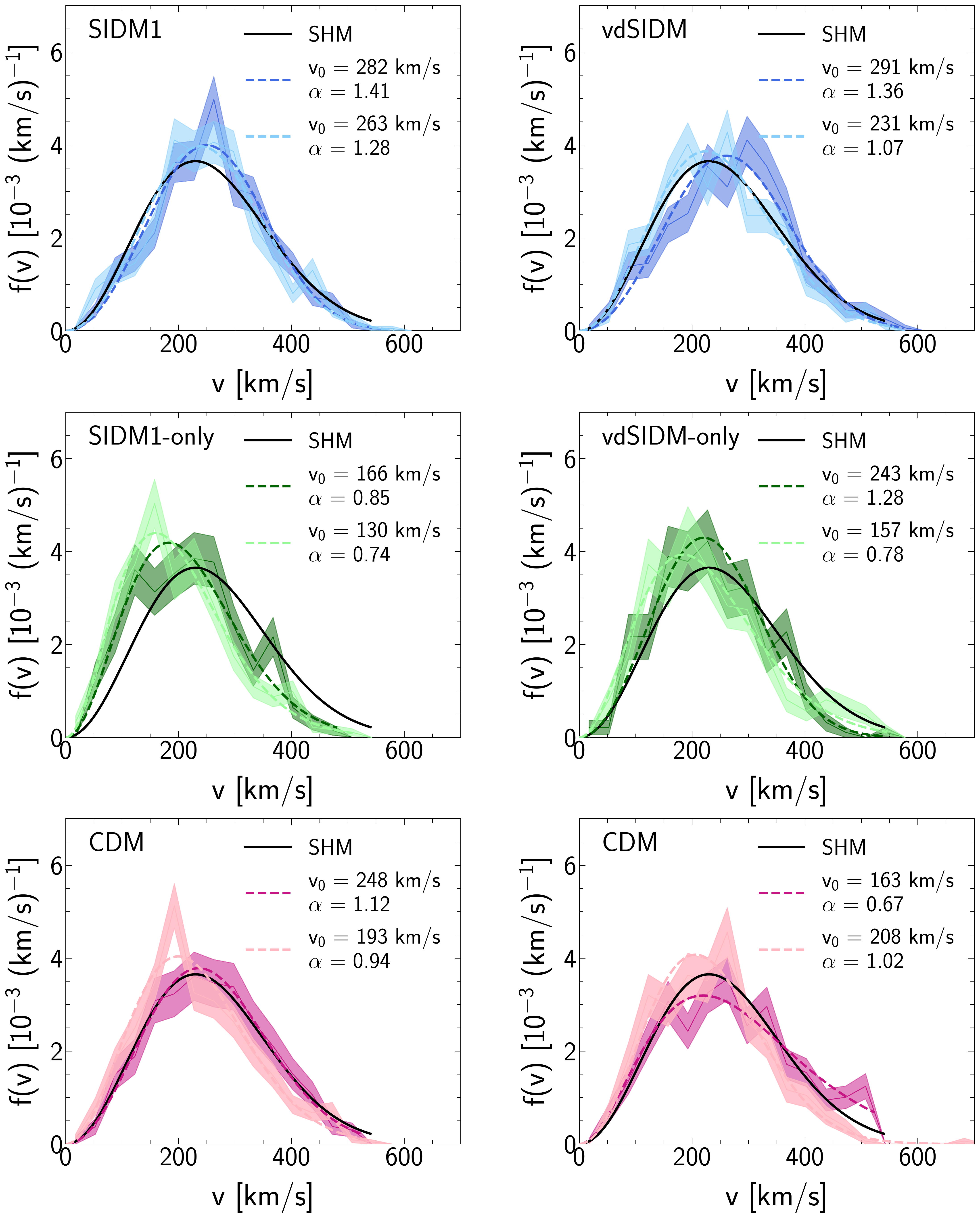}
  \caption{The local DM velocity modulus distributions (solid coloured curves) and their 1$\sigma$ Poisson uncertainties (shaded bands) in the Galactic rest frame for the SIDM1 and vdSIDM  halos (top panels) that have the best (dark colour) and worst (light colour) fit to the truncated generalized Maxwellian velocity distribution (eq.~\eqref{eq:genmax}). The middle and bottom panels represent the SIDM-only and CDM counterparts of the SIDM halos, respectively. The best fit truncated generalized Maxwellian distribution for each halo is plotted as the dashed curve, and the SHM Maxwellian distribution with a peak speed of 230 km$/$s is shown as the black solid curve. For each halo the values of the best fit parameters of the generalized Maxwellian distribution, $v_0$ and $\alpha$, are indicated on the panels. }
  \label{fig: best_worst_vel_dist_genMax}
\end{figure}

In figure~\ref{fig: best_worst_vel_dist_genMax}, we show the local DM velocity modulus distributions and their 1$\sigma$ uncertainties for the SIDM1 and vdSIDM halos (top panels) that have the best (dark colour) and worst (light colour) fits to the truncated generalized Maxwellian distribution (eq.~\eqref{eq:genmax}). The corresponding SIDM-only and CDM halos are shown in the middle and bottom panels, respectively. The dashed lines indicate the best fit generalized Maxwellian distributions, with the best fit parameters quoted for each halo. The solid black line shows the SHM Maxwellian distribution with a peak speed of 230~km/s.

\subsection{Best fit parameters for velocity distributions}\label{app:parameters}

In tables~\ref{tab:eagle-50-sidm1}, \ref{tab:eagle-50-sidm1-DMO}, and \ref{tab:eagle-50-sidm1-CDM} we quote the best fit parameters for the truncated Maxwellian (eq.~\eqref{eq:maxwellian}) and the generalized Maxwellian (eq.~\eqref{eq:genmax}) speed distributions to fit the DM velocity modulus distributions for the SIDM1 halos, as well as their SIDM1-only, and CDM counterparts, respectively. Similarly, the best fit parameters for the vdSIDM halos and their vdSIDM-only and CDM counterparts are quoted in tables~\ref{tab:eagle-50-vdsidm}, \ref{tab:eagle-50-vdsidm-DMO}, and \ref{tab:eagle-50-vdsidm-CDM}, respectively.

In tables~\ref{tab:eagle-50-comp-sidm1}, \ref{tab:eagle-50-comp-sidm1-DMO}, and \ref{tab:eagle-50-comp-sidm1-CDM}, we present the best fit parameters for the Gaussian distribution (eq.~\eqref{eq:gaussian}) to fit the radial, azimuthal, and vertical components of the local DM velocity distributions of the SIDM1 halos, and their SIDM1-only and CDM counterparts, respectively. Similarly the best fit parameters for the Gaussian fit for the vdSIDM halos and their vdSIDM-only and CDM counterparts are listed in tables~\ref{tab:eagle-50-comp-vdsidm}, \ref{tab:eagle-50-comp-vdsidm-DMO}, and \ref{tab:eagle-50-comp-vdsidm-CDM}, respectively.

\begin{table}[ht]
\centering
\begin{tabular}{|c|*{5}{w{c}{5em}|}}
 \hline
 \multicolumn{6}{|c|}{SIDM1} \\
    \hline
      & \multicolumn{2}{c|}{Maxwellian}        
                                & \multicolumn{3}{c|}{Generalized Maxwellian}                                  \\ 
      \hline
Halo Number &     $v_0$~[km/s]   &   $\chi^2/(N-1)$    &   $v_0$~[km/s]   & $\alpha$   & $\chi^2/(N-2)$  \\ \hline
48     &     238   &   1.27           &   244       &   1.03   &   1.34   \\ \hline
60     &     243   &   1.25           &   266       &   1.12   &   1.17   \\ \hline
71     &     232   &   0.68           &   216       &   0.92   &   0.66   \\ \hline
74     &     242   &   1.22           &   258       &   1.09   &   1.20   \\ \hline
80     &     245   &   1.39           &   273       &   1.15   &   1.32   \\ \hline
92     &     246   &   0.77           &   270       &   1.13   &   0.64   \\ \hline
96     &     224   &   1.36           &   229       &   1.03   &   1.45   \\ \hline
99     &     233   &   1.78           &   287       &   1.36   &   0.83   \\ \hline
102     &    224   &   2.43           &   282       &   1.41   &   1.12   \\ \hline
111     &    219   &   0.95           &   253       &   1.23   &   0.67   \\ \hline
118     &    221   &   2.52           &   263       &   1.28   &   2.02   \\ \hline
128     &    243   &   0.83           &   239       &   0.98   &   0.88  \\ \hline
131     &    241   &   0.47           &   251       &   1.05   &   0.48   \\ \hline
159     &    218   &   0.76           &   225       &   1.04   &   0.79   \\ \hline
\end{tabular}
\captionof{table}{\label{tab:eagle-50-sidm1} Best fit parameters for the standard and generalized Maxwellian distributions (eqs.~\eqref{eq:maxwellian} and \eqref{eq:genmax}) to fit the local DM velocity modulus distributions of the  MW-like SIDM1 halos. The reduced $\chi^2$ values, $\chi^2/(N- dof)$, are also quoted in the table.}
    \end{table}

\begin{table}[ht]
\centering
\begin{tabular}{|c|*{5}{w{c}{5em}|}}
 \hline
 \multicolumn{6}{|c|}{SIDM1-only counterparts} \\
    \hline
      & \multicolumn{2}{c|}{Maxwellian}        
                                & \multicolumn{3}{c|}{Generalized Maxwellian}                                  \\ 
      \hline
Halo Number &     $v_0$~[km/s]    &   $\chi^2/(N-1)$    &   $v_0$~[km/s]    & $\alpha$   & $\chi^2/(N-2)$  \\ \hline
49     &     194   &   1.21           &   161       &   0.82   &   1.01   \\ \hline
64     &     214   &   0.53          &   223       &   1.05   &   0.56   \\ \hline
74     &     191   &   2.51           &   128       &   0.68   &   1.56   \\ \hline
78     &     202   &   0.83           &   171       &   0.83   &   0.72   \\ \hline
84     &     204   &   0.47           &   182       &   0.88   &   0.40   \\ \hline
94     &     212   &   1.17           &   191       &   0.89   &   1.17   \\ \hline
104     &    187   &   1.05           &   138       &   0.74   &   0.55   \\ \hline
107     &    201   &   1.47           &   163       &   0.80   &   1.02   \\ \hline
119     &    194   &   1.37           &   166       &   0.85   &   1.35   \\ \hline
121     &    181   &   1.76           &   154       &   0.85   &   1.60   \\ \hline
136     &    176   &   1.87           &   130       &   0.74   &   1.00   \\ \hline
127     &    217   &   0.91           &   206       &   0.94   &   0.96  \\ \hline
130     &    219   &   1.09           &   177   &   0.80   &   0.77   \\ \hline
159     &    186   &   0.97          &   163       &   0.86   &   0.85   \\ \hline
\end{tabular}
\captionof{table}{\label{tab:eagle-50-sidm1-DMO} Same as table~\ref{tab:eagle-50-sidm1}, but for the corresponding SIDM1-only halos. The order of the halo numbers is the same as their matched SIDM1 counterparts in table~\ref{tab:eagle-50-sidm1}.}
    \end{table}

\begin{table}[ht]
\centering
\begin{tabular}{|c|*{5}{w{c}{5em}|}}
 \hline
 \multicolumn{6}{|c|}{CDM counterparts of SIDM1 halos} \\
    \hline
      & \multicolumn{2}{c|}{Maxwellian}        
                                & \multicolumn{3}{c|}{Generalized Maxwellian}                                  \\ 
      \hline
Halo Number &     $v_0$~[km/s]    &   $\chi^2/(N-1)$    &   $v_0$~[km/s]    & $\alpha$   & $\chi^2/(N-2)$  \\ \hline
47     &     241   &   1.59           &   163       &   0.67   &   1.07   \\ \hline
68     &     224   &   1.19           &   216       &   0.96   &   1.28   \\ \hline
77     &     205   &   1.00           &   208       &   1.02   &   1.08   \\ \hline
74     &     236   &   1.06           &   234       &   0.99   &   1.15   \\ \hline
83     &     213   &   0.75           &   199       &   0.93   &   0.75  \\ \hline
93     &     236   &   1.56           &   268       &   1.17   &   1.45   \\ \hline
96     &     222   &   0.44           &   222       &   1.00   &   0.48   \\ \hline
98     &     230   &   1.20           &   266       &   1.20   &   0.82   \\ \hline
114     &    228   &   0.33           &   248       &   1.12   &   0.28   \\ \hline
110     &    206   &   1.35           &   224       &   1.11   &   1.36   \\ \hline
123     &    204   &   1.33           &   193       &   0.94   &   1.37   \\ \hline
146     &    215   &   1.07           &   204       &   0.94  &   1.14  \\ \hline
136     &    214   &   0.75           &   218       &   1.02   &   0.79   \\ \hline
159     &    213   &   0.92           &   215       &   1.01   &   1.00   \\ \hline
\end{tabular}
\captionof{table}{\label{tab:eagle-50-sidm1-CDM} Same as table~\ref{tab:eagle-50-sidm1}, but for the corresponding CDM halos. The order of the halo numbers is the same as their matched SIDM1 counterparts in table~\ref{tab:eagle-50-sidm1}.}
    \end{table}

\begin{table}[ht]
\centering
\begin{tabular}{|c|*{16}{w{c}{5em}|}}
 \hline
 \multicolumn{6}{|c|}{vdSIDM} \\
    \hline
      & \multicolumn{2}{c|}{Maxwellian}        
                                & \multicolumn{3}{c|}{Generalized Maxwellian}                                  \\ 
      \hline
Halo Number &     $v_0$~[km/s]    &   $\chi^2/(N-1)$    &   $v_0$~[km/s]    & $\alpha$   & $\chi^2/N-2$  \\ \hline
55  &   237  &  1.47   &   291     &   1.36     &   1.03   \\ \hline
60  &   246  &  1.11   &   237     &   0.96     &   1.17   \\ \hline
72  &   220  &  1.59   &   231     &   1.07     &   1.68   \\ \hline
74  &   237  &  0.84   &   267     &   1.17     &   0.63   \\ \hline
75  &   254  &  1.61   &   269     &   1.08     &   1.67   \\ \hline
77  &   214  &  0.72   &   219     &   1.03     &   0.77   \\ \hline
92  &   258  &  0.69   &   250     &   0.96     &   0.72   \\ \hline
93  &   230  &  2.61   &   299     &   1.44     &   1.14   \\ \hline
99  &   228  &  1.96   &   266     &   1.24     &   1.53   \\ \hline
101 &   245  &  0.34   &   245     &   1.00     &   0.36   \\ \hline
107 &   226  &  1.31   &   261     &   1.21     &   0.85   \\ \hline
115 &   221  &  1.53   &   262     &   1.27     &   1.17   \\ \hline
122 &   246  &  0.82   &   238     &   0.96     &   0.90   \\ \hline
124 &   235  &  0.74   &   251     &   1.09     &   0.71   \\ \hline
128 &   205  &  0.93   &   225     &   1.12     &   0.96   \\ \hline
138 &   229  &  1.70   &   277     &   1.29     &   1.18   \\ \hline
176 &   224  &  0.70   &   235     &   1.06     &   0.73   \\ \hline
                                        
\end{tabular}
\captionof{table}{\label{tab:eagle-50-vdsidm} Best fit parameters for the standard and generalized Maxwellian distributions to fit the local DM velocity modulus distributions of the  MW-like vdSIDM halos.}
\end{table}

\begin{table}[ht]
\centering
\begin{tabular}{|c|*{16}{w{c}{5em}|}}
 \hline
 \multicolumn{6}{|c|}{vdSIDM-only counterparts} \\
    \hline
      & \multicolumn{2}{c|}{Maxwellian}        
                                & \multicolumn{3}{c|}{Generalized Maxwellian}                                  \\ 
      \hline
Halo Number &     $v_0$~[km/s]    &   $\chi^2/(N-1)$    &  $v_0$~[km/s]   & $\alpha$   & $\chi^2/(N-2)$  \\ \hline
47  &   205  &  1.06   &   243     &   1.28     &   1.20   \\ \hline
64  &   213  &  0.85   &   195     &   0.91     &   0.88   \\ \hline
73  &   196  &  1.29   &   157     &   0.78    &   1.08   \\ \hline
77  &   196  &  0.86   &   204     &   1.06     &   0.93   \\ \hline
83  &   206  &  1.05   &   177     &   0.84     &   1.00   \\ \hline
72  &   203  &  1.16   &   182     &   0.88     &   1.12   \\ \hline
74  &   220  &  2.61   &   141     &   0.68     &   1.64   \\ \hline
108  &   210  &  0.66   &   213     &   1.02     &   0.71   \\ \hline
117  &   196 &  0.59   &   194     &   0.99     &   0.64   \\ \hline
85 &   231  &  1.36   &   220     &   0.94     &    1.46   \\ \hline
111 &   195  &  0.62   &   168     &   0.85     &   0.53   \\ \hline
119 &   182  &  1.11   &   204     &   1.15     &   1.06   \\ \hline
105 &   238  &  0.62   &   234     &   0.98     &   0.68   \\ \hline
131 &   212  &  0.54   &   225     &   1.08     &   0.55   \\ \hline
128 &   158  &  2.12   &   175     &   1.15     &   2.28   \\ \hline
127 &   211  &  0.66   &   202     &   0.95     &   0.71   \\ \hline
162 &   211  &  0.97   &   184     &   0.85     &   0.96   \\ \hline
                                        
\end{tabular}
\captionof{table}{\label{tab:eagle-50-vdsidm-DMO} Same as table~\ref{tab:eagle-50-vdsidm}, but for the corresponding vdSIDM-only halos. The order of the halo numbers is the same  as their matched vdSIDM counterparts in table~\ref{tab:eagle-50-vdsidm}.}
    \end{table}

\begin{table}[ht]
\centering
\begin{tabular}{|c|*{16}{w{c}{5em}|}}
 \hline
 \multicolumn{6}{|c|}{CDM counterparts of vdSIDM halos} \\
    \hline
      & \multicolumn{2}{c|}{Maxwellian}        
                                & \multicolumn{3}{c|}{Generalized Maxwellian}                                  \\ 
      \hline
Halo Number &     $v_0$~[km/s]    &   $\chi^2/(N-1)$    &   $v_0$~[km/s]    & $\alpha$   & $\chi^2/(N-2)$  \\ \hline
47  &   241  &  1.59   &   163     &   0.67     &   1.07   \\ \hline
68  &   224  &  1.19   &   216     &   0.96     &   1.28   \\ \hline
77  &   205  &  1.00   &   208     &   1.02     &   1.08   \\ \hline
74  &   237  &  1.06   &   234     &   0.99     &   1.15   \\ \hline
83  &   213  &  0.75   &   199     &   0.93     &   0.75  \\ \hline
75  &   220  &  2.20   &   223     &   1.02     &   2.38   \\ \hline
87  &   241  &  0.69   &   266     &   1.14     &   0.61   \\ \hline
98  &   230  &  1.20   &   266     &   1.20     &   0.82   \\ \hline
114  &   228  &  0.33   &   248     &   1.12     &   0.28   \\ \hline
89 &   231  &  1.28   &   208     &   0.89     &   1.25   \\ \hline
115 &   210  &  2.04   &   217     &   1.04     &   2.21   \\ \hline
110 &   206  &  1.36   &   224     &   1.11     &   1.36   \\ \hline
116 &   213  &  1.18   &   177     &   0.83     &   0.75   \\ \hline
136 &   214  &  0.75   &   218     &   1.02     &   0.79   \\ \hline
149 &   169  &  1.21   &   138     &   0.81     &   0.92   \\ \hline
146 &   215  &  1.08   &   204     &   0.94     &   1.14   \\ \hline
182 &   213  &  1.91   &   198     &   0.93     &   2.08   \\ \hline
\end{tabular}
\captionof{table}{\label{tab:eagle-50-vdsidm-CDM} Same as table~\ref{tab:eagle-50-vdsidm}, but for the corresponding CDM halos. The order of the halo numbers is the same as their matched vdSIDM counterparts in table~\ref{tab:eagle-50-vdsidm}.}
\end{table}


\begin{sidewaystable}
\centering
\begin{tabular}{|c|c|c|c|c|c|c|c|c|c|} 
 \hline
 \multicolumn{10}{|c|}{SIDM1} \\
\hline
Halo Number & $v_{0, \rho}$~[km/s] & $\mu_{\rho}$~[km/s] & $\chi_{\rho}^2/(N-2)$ & $v_{0, \phi}$~[km/s] & $\mu_{0, \phi}$~[km/s] & $\chi_{\phi}^2/(N-2)$ & $v_{0, z}$~[km/s] & $\mu_{z}$~[km/s] & $\chi_{z}^2/(N-2)$  \\ 
\hline
48     & 242    & -2.43     & 0.619      & 232      & 39.4       & 0.964        & 216  & -24.8   & 1.33     \\ 
\hline
60     & 242    & -4.82     & 0.831      & 244      & 25.6       & 1.00        & 244  & 33.6   & 1.05     \\ 
\hline
71     & 228    & -4.02     & 1.13      & 221      & 16.2       & 1.09        & 224  & 12.8   & 0.868     \\
\hline
74     & 241    & -0.802     & 0.607      & 255      & 1.00       & 0.612        & 230  & -17.6   & 0.602     \\
\hline
80     & 236    & -12.6     & 0.377      & 239      & 17.2       & 0.775        & 236  & 34.4   & 0.622     \\
\hline
92     & 237    & -10.4     & 0.476      & 247      & 22.6       & 0.707        & 250  & -6.4   & 0.674     \\
\hline
96     & 216    & 8.81     & 0.621      & 230      & 27.4       & 0.945        & 216  & 14.4   & 1.10     \\
\hline
99     & 237    & -3.20     & 0.664      & 220      & 71.8       & 1.25        & 236  & 7.22   & 0.751     \\
\hline
102     & 236    & 4.00     & 0.876      & 213      & 5.81       & 1.28        & 229  & 3.20   & 1.06     \\
\hline
111     & 207    & 13.6     & 0.656      & 221      & 64.0       & 0.895        & 194  & 9.62   & 1.228     \\
\hline
118     & 221    & -1.61     & 1.23      & 204      & 47.8       & 1.17        & 216  & -2.40   & 1.03     \\
\hline
128     & 257    & -1.60     & 0.951      & 241      & 9.43       & 0.666        & 214  & 4.82   & 1.042     \\
\hline
131     & 237    & -7.28     & 0.663      & 250      & 21.4       & 0.669        & 222  & -3.21   & 1.48     \\
\hline
159     & 212    & 20.0     & 0.968      & 201      & 32.2       & 0.697        & 223  & 12.0   & 0.647     \\
\hline
\end{tabular}
\captionof{table}{\label{tab:eagle-50-comp-sidm1} Best fit parameters for the Gaussian distribution (eq.~\eqref{eq:gaussian}) to fit the radial, azimuthal, and vertical components of the local DM velocity distributions of the  MW-like SIDM1 halos.}
\end{sidewaystable}

\begin{sidewaystable}
\centering
\begin{tabular}{|c|c|c|c|c|c|c|c|c|c|} 
 \hline
 \multicolumn{10}{|c|}{SIDM1-only counterparts} \\
\hline
Halo Number & $v_{0, \rho}$~[km/s] & $\mu_{\rho}$~[km/s] & $\chi_{\rho}^2/(N-2)$ & $v_{0,\phi}$~[km/s] & $\mu_{\phi}$~[km/s] & $\chi_{\phi}^2/(N-2)$ & $v_{0,z}$~[km/s] & $\mu_{z}$~[km/s] & $\chi_{z}^2/(N-2)$  \\ 
\hline
49     & 197    & 13.6     & 1.03      & 189      & 8.81       & 0.712        & 186  & -24.0   & 0.772     \\ 
\hline
64     & 218    & 22.4     & 0.869      & 210      & 18.4       & 0.601        & 197  & 24.0   & 2.46     \\ 
\hline
74     & 175    & 3.2     & 0.869      & 187      & -8.65       & 0.977        & 200  & 3.21   & 0.726     \\
\hline
78     & 185    & 0.00     & 0.741      & 204      & 4.65       & 0.628        & 189  & -32.8   & 0.886     \\
\hline
84     & 198    & -4.86     & 0.761      & 194      & 8.46       & 1.26       & 185  & 15.2   & 1.24     \\
\hline
94     & 212    & 20.0     & 0.943      & 193      & -25.4       & 1.58        & 210  & 7.22   & 0.772     \\
\hline
104     & 205    & 4.82     & 0.555      & 180      & -7.42       & 1.01        & 171  & 13.6   & 0.852     \\
\hline
107     & 214    & -6.4     & 0.548      & 192      & 16.0       & 0.754        & 195  & -9.62   & 0.781     \\
\hline
119     & 191    & -8.08     & 1.04      & 186      & 0.42      & 0.653        & 196  & -4.82   & 0.886     \\
\hline
121     & 187    & -6.42     & 0.857      & 188      & 25.6       & 0.814        & 150  & 8.04   & 0.496     \\
\hline
136     & 180    & -7.21     & 1.08      & 164      & 32.2       & 1.29        & 156  & -3.22   & 1.18     \\
\hline
127     & 226    & 0.802     & 0.921      & 215      & 5.82       & 0.733       & 195  & -4.03   & 1.17     \\
\hline
130     & 223    & -9.60     & 0.603      & 214      & -3.82       & 0.664        & 203  & 6.40   & 0.467     \\
\hline
159     & 190    & -17.6     & 0.743      & 179      & -0.20       & 0.431        & 180  & 4.82   & 1.14     \\
\hline
\end{tabular}
\captionof{table}{\label{tab:eagle-50-comp-sidm1-DMO} Same as table~\ref{tab:eagle-50-comp-sidm1}, but for the corresponding SIDM1-only halos.}
\end{sidewaystable}

\begin{sidewaystable}
\centering
\begin{tabular}{|c|c|c|c|c|c|c|c|c|c|} 
\hline
 \multicolumn{10}{|c|}{CDM counterparts of SIDM1 halos} \\
\hline
Halo Number & $v_{0,\rho}$~[km/s] & $\mu_{\rho}$~[km/s] & $\chi_{\rho}^2/(N-2)$ & $v_{0,\phi}$~[km/s] & $\mu_{\phi}$~[km/s] & $\chi_{\phi}^2/(N-2)$ & $v_{0,z}$~[km/s] & $\mu_{z}$~[km/s] & $\chi_{z}^2/(N-2)$  \\ 
\hline
47     & 235    & 7.22     & 1.05      & 224      & 21.4       & 2.79        & 206  & -43.2   & 1.09     \\ 
\hline
68     & 224    & -14.4     & 1.13      & 224      & 38.8       & 0.774        & 193  & 6.42   & 1.43     \\ 
\hline
74     & 272    & 7.2     & 0.898      & 238      & -5.61       & 1.82        & 205  & -12.8   & 1.00     \\
\hline
77     & 213    & 0.807     & 1.09      & 211      & -2.20       & 0.813        & 197  & 12.3   & 0.515     \\
\hline
83     & 234    & 0.00     & 0.814      & 212      & 23.8       & 1.03        & 187  & 2.43   & 0.732     \\
\hline
93     & 257    & 1.61     & 0.841      & 240      & 30.2       & 0.635        & 206  & 1.62   & 2.07     \\
\hline
96     & 212    & -7.26     & 1.68      & 233      & 19.6       & 0.958        & 201  & 12.0   & 1.82     \\
\hline
98     & 215    & -7.21     & 0.872      & 223      & 79.6       & 1.05        & 217  & 9.61   & 0.819     \\
\hline
110     & 218    & 2.42     & 0.749      & 179      & 79.6      & 2.14        & 178  & -12.0   & 0.812     \\
\hline
114     & 235    & 12.0     & 0.887      & 194      & 6.41       & 1.07        & 205  & 79.2   & 1.02     \\
\hline
123     & 196    & 2.41     & 0.907      & 195      & 79.0       & 1.49        & 178  & 4.01   & 1.375     \\
\hline
136     & 238    & 2.41     & 1.29      & 192      & 14.8       & 1.14       & 199  & 0.806   & 0.657     \\
\hline
146     & 225    & -0.827     & 0.555      & 210      & -5.65       & 1.23        & 190  & -5.66   & 0.605     \\
\hline
159     & 232    & -7.22     & 0.587      & 196      & 65.2       & 1.02        & 183  & 9.65   & 1.47     \\
\hline
\end{tabular}
\captionof{table}{\label{tab:eagle-50-comp-sidm1-CDM} Same as table~\ref{tab:eagle-50-comp-sidm1}, but for the corresponding CDM halos.}
\end{sidewaystable}

\begin{sidewaystable}
\centering
\begin{tabular}{|c|c|c|c|c|c|c|c|c|c|} 
\hline
 \multicolumn{10}{|c|}{vdSIDM} \\
\hline
Halo Number & $v_{0,\rho}$~[km/s] & $\mu_{\rho}$~[km/s] & $\chi_{\rho}^2/(N-2)$ & $v_{0,\phi}$~[km/s] & $\mu_{\phi}$~[km/s] & $\chi_{\phi}^2/(N-2)$ & $v_{0,z}$~[km/s] & $\mu_{z}$~[km/s] & $\chi_{z}^2/(N-2)$  \\ 
\hline
55     & 223    & -4.01     & 0.565      & 249      & 44.2       & 0.728        & 233  & -24.8   & 1.16     \\ 
\hline
60     & 251    & -12.0     & 0.730      & 233      & 14.8       & 0.995        & 241  & 0.801   & 1.14     \\ 
\hline
72     & 215    & -4.01     & 0.661      & 217      & -7.41       & 0.818        & 214  & 8.11   & 1.53     \\
\hline
74     & 239    & -0.01    & 0.766      & 231      & -12.8      & 1.25        & 222  & -8.02   & 1.323     \\
\hline
75     & 247    & 5.62      & 0.785      & 242      & -10.4      & 1.19        & 266  & 20.8   & 1.17     \\
\hline
77     & 209    & 2.39     &  1.12      & 205      & 36.4       & 1.28        & 193  & 12.8   & 0.934     \\
\hline
92     & 249    & 4.19     & 1.06      & 240      & 10.6       & 0.733        & 237  & -7.19   & 0.977     \\
\hline
93     & 226    & -0.11     & 1.28       & 238      & 77.8       & 1.15        & 225  & -0.01   & 0.842     \\
\hline
99     & 224    & 1.61     & 0.541       & 221      & 10.0        & 1.08        & 217  & 4.88   & 1.47     \\
\hline
101     & 237    & 20.8     & 0.927     & 231      & 7.63        & 1.00        & 239  & -0.80   & 1.31     \\
\hline
107     & 231    & 4.78     & 0.814      & 222      & 4.59        & 0.816        & 225  & 13.6   & 0.755     \\
\hline
115     & 220    & -3.21     & 1.36     & 220      & 48.4       & 0.734        & 213  & 5.64   & 1.31     \\
\hline
122     & 261    & -12.0    & 1.13    & 229      & 11.8        & 0.673        & 205  & 5.61   & 0.979     \\
\hline
124     & 223    & 2.43     & 1.17      & 246      & 14.2       & 0.867        & 226  & 3.21   & 0.841     \\
\hline
128     & 223    & 1.61     & 1.49      & 208       & 9.41      & 1.35        & 188  & 6.44   & 1.38     \\
\hline
138     & 230    & -7.21    & 0.495      & 229       & 22.6     & 0.952        & 237  & -3.21  & 0.831     \\
\hline
176     & 222    & -12.8   & 1.30      & 227       & 7.00       & 0.471        & 221  & 8.81  & 0.874     \\
\hline

\end{tabular}
\captionof{table}{\label{tab:eagle-50-comp-vdsidm} Best fit parameters for the Gaussian distribution to fit the radial, azimuthal, and vertical components of the local DM velocity distributions of the  MW-like vdSIDM halos.}
\end{sidewaystable}

\begin{sidewaystable}
\centering
\begin{tabular}{|c|c|c|c|c|c|c|c|c|c|} 
\hline
 \multicolumn{10}{|c|}{vdSIDM-only counterparts} \\
\hline
Halo Number & $v_{0,\rho}$~[km/s] & $\mu_{\rho}$~[km/s] & $\chi_{\rho}^2/(N-2)$ & $v_{0,\phi}$~[km/s] & $\mu_{\phi}$~[km/s] & $\chi_{\phi}^2/(N-2)$ & $v_{0,z}$~[km/s] & $\mu_{z}$~[km/s] & $\chi_{z}^2/(N-2)$  \\ 
\hline
47     & 207    & -8.45     & 1.48      & 220      & -14.0       & 1.31     & 203  & -15.2   & 0.404     \\ 
\hline
64     & 212    & 10.4     & 1.40      & 213      & 8.21       & 0.687       & 194  & -21.6   & 1.61     \\ 
\hline
73     & 193    & 2.43     & 2.11      & 187      & 9.45       & 0.919       & 188  & 3.26   & 0.997     \\
\hline
77     & 197    & -14.4     & 0.766      & 206      & 21.4       & 1.15     & 168  & 20.0   & 0.779     \\
\hline
83     & 205    & -1.62     & 0.866      & 193      & -21.8       & 0.942     & 198  & 21.6   & 0.884     \\
\hline
72     & 222    & 5.65     & 0.405      & 194      & 16.6       & 0.657       & 180  & -3.25   & 0.485     \\
\hline
74     & 227    & -5.67     & 0.804      & 216      & -0.211       & 0.695      & 211  & -11.2   & 0.906     \\
\hline
108     & 207    & -0.812     & 1.02      & 214      & 29.2       & 1.61      & 188  & 10.4   & 0.803     \\
\hline
117     & 181    & -12.8     & 0.721      & 190      & -9.22      & 0.907     & 217  & -16.0   & 0.832     \\
\hline
85     & 249    & 7.22     & 1.11      & 197      & 2.25       & 1.29        & 220  & -3.22   & 0.835     \\
\hline
111     & 196    & -0.899     & 0.836      & 186      & -8.27       & 1.67    & 183  & -4.25   & 1.13     \\
\hline
119     & 195    & -1.67     & 0.453      & 183      & 39.7       & 1.08     & 183  & 1.59   & 0.767     \\
\hline
105     & 245    & 0.878     & 0.727      & 234      & 5.2       & 0.857     & 240  & -1.63   & 1.09     \\
\hline
131     & 219    & -4.83     & 0.795      & 202      & 14.2       & 1.09      & 209  & -1.64   & 0.652     \\
\hline
128     & 155    & 2.45      & 1.05      & 170      & 7.62     & 0.744       & 148  & 28.2   & 0.622     \\
\hline
127     & 206    & -7.2    & 1.077      & 235      & -0.20     & 0.824       & 187  & -4.02   & 1.07     \\
\hline
162     & 227    & -2.43      & 0.721      & 205      & 4.00      & 1.01       & 194  & 6.38    & 1.06     \\
\hline
\end{tabular}
\captionof{table}{\label{tab:eagle-50-comp-vdsidm-DMO} Same as table~\ref{tab:eagle-50-comp-vdsidm}, but for the corresponding vdSIDM-only halos.}
\end{sidewaystable}

\begin{sidewaystable}
\centering
\begin{tabular}{|c|c|c|c|c|c|c|c|c|c|} 
\hline
 \multicolumn{10}{|c|}{CDM counterparts of vdSIDM halos} \\
\hline
Halo Number & $v_{0,\rho}$~[km/s] & $\mu_{\rho}$~[km/s] & $\chi_{\rho}^2/(N-2)$ & $v_{0,\phi}$~[km/s] & $\mu_{\phi}$~[km/s] & $\chi_{\phi}^2/(N-2)$ & $v_{0,z}$~[km/s] & $\mu_{z}$~[km/s] & $\chi_{z}^2/(N-2)$  \\  
\hline
47     & 235    & 7.21    & 1.05      & 224      & 21.4       & 2.79       & 206  & -43.2   & 1.09     \\ 
\hline
68     & 224    & -14.4     & 1.13      & 224      & 38.8       & 0.774    & 193  & 6.42   & 1.43     \\ 
\hline
77     & 213    & 0.811     & 1.09      & 211      & -2.02       & 0.813      & 197  & 12.3   & 0.515     \\
\hline
74     & 272    & 7.24     & 0.898      & 238      & -5.64      & 1.82       & 205  & -12.8   & 1.01     \\
\hline
83     & 234    & 0.008     & 0.814      & 212      & 23.8       & 1.03      & 187  & 2.44   & 0.732     \\
\hline
75     & 265    & -16.2     & 0.996      & 198      & 32.8       & 1.29    & 214  & -12.8   & 1.15     \\
\hline
87     & 244    & 9.69     & 0.959      & 203      & -22.4       & 0.708     & 265  & 8.05   & 0.757     \\
\hline
98     & 215    & -7.22     & 0.872      & 223      & 79.6       & 1.05     & 217  & 9.65   & 0.819     \\
\hline
114     & 235    & 11.9     & 0.887      & 194      & 6.44      & 1.07      & 205  & 79.2   & 1.02     \\
\hline
89     & 240    & -8.10     & 0.768      & 222      & 53.2       & 1.05     & 203  & 6.45   & 0.965     \\
\hline
115     & 235    & -3.23     & 1.30      & 207      & 2.22       & 1.13     & 180  & 0.00   & 0.604     \\
\hline
110     & 218    & 2.41     & 0.749      & 179      & 79.6       & 2.14     & 178  & -12.0   & 0.812     \\
\hline
116     & 235    & -4.09     & 0.967      & 215      & 0.403       & 1.47     & 179  & 1.61   & 0.885     \\
\hline
136     & 238    & 2.42     & 1.29      & 192      & 14.8       & 1.14    & 199  & 0.811   & 0.657     \\
\hline
149     & 188    & -6.42      & 0.853     & 162       & 3.42      & 1.16      & 146  & 12.1  & 0.664      \\
\hline
146     & 225    & -0.822     & 0.555     & 210       & -5.68    & 1.23      & 190  & -5.62  & 0.605      \\
\hline
182     & 237   & 2.43    & 0.961      & 186       & -17.3    & 1.34      & 198  & -4.00  & 1.27      \\
\hline

\end{tabular}
\captionof{table}{\label{tab:eagle-50-comp-vdsidm-CDM} Same as table~\ref{tab:eagle-50-comp-vdsidm}, but for the corresponding CDM halos.}
\end{sidewaystable}

\bibliographystyle{JHEP}
\bibliography{./refs}

\end{document}